\documentclass[11pt]{report}
\usepackage{authblk}
\usepackage{booktabs}
\usepackage[utf8]{inputenc}
\usepackage{amsmath}
\usepackage{amsfonts}
\usepackage{amssymb}
\usepackage{graphicx}
\usepackage[left=2cm,right=2cm,top=2cm,bottom=2cm]{geometry}

\usepackage{bera}
\usepackage{libertine}
\usepackage[libertine]{newtxmath}

\usepackage[ruled]{algorithm2e} % For algorithms

\SetAlFnt{\small}
\SetAlCapFnt{\small}
\SetAlCapNameFnt{\small}
\SetAlCapHSkip{0pt}
\IncMargin{-\parindent}
\usepackage{tikz}
\usepackage{subfig}
\usepackage{booktabs} % To thicken table lines
\usetikzlibrary{quotes,arrows.meta}
\usepackage{tikz-dimline}
\pgfplotsset{compat=newest}
\usepackage{mathdots}
\usepackage{svg}
\usepackage{hyperref}
\usepackage{lscape}
\usepackage{bm}
\usepackage[mathscr]{euscript}
\usepackage{forest,array}
\usepackage{adjustbox}
\usepackage[nolist]{acronym}
  \newacro{ml}[ML]{Machine Learning}
  \newacro{ai}[AI]{Artificial Intelligence}
  \newacro{dl}[DL]{Deep Learning}
  \newacro{mlp}[MLP]{Multi-Layer Perceptron}
  \newacro{cnn}[CNN]{Convolutional Neural Network}
  \newacro{dnn}[DNN]{Deep Neural Network}
  \newacro{bnn}[BNN]{Binary Neural Network}
  \newacro{fm}[FM]{Feature Map}
  \newacro{relu}[ReLU]{Rectified Linear Unit}
  \newacro{mnist}[MNIST]{Modified National Institute of Standards and Technology}
  \newacro{ilsvrc}[ILSVRC]{ImageNet Large Scale Visual Recognition Competition}
  \newacro{asic}[ASIC]{Application Specific Integrated Circuits}
  \newacro{dram}[DRAM]{Dynamic Random Access Memory}
  \newacro{fpga}[FPGA]{Field-Programmable Gate Array}
  \newacro{gpu}[GPU]{Graphics Processing Unit}
  \newacro{gpp}[GPP]{General Purpose Processor}
  \newacro{cpu}[CPU]{Central Processing Unit}
  \newacro{pe}[PE]{Processing Element}
  \newacro{simd}[SIMD]{Single Instruction on Multiple Data}
  \newacro{simt}[SIMD]{Single Instruction on Multiple Threads}  
  \newacro{gemm}[GEMM]{General Matrix Multiplication}
  \newacro{ewmm}[EWMM]{Element-Wise Matrix Multiplication}
  \newacro{fft}[FFT]{Fast Fourier Transofrm}
  \newacro{dsp}[DSP]{Digital Signal Processing}
  \newacro{hdl}[HDL]{Hardware Description Language}
  \newacro{le}[LE]{Logic Elements}
  \newacro{fifo}[FIFO]{First-In First-Out}
  \newacro{haddoc}[HADDOC]{Hardware Automated Dataflow Description Of CNNs}
  \newacro{hls}[HLS]{High-Level Synthesis}
  \newacro{hpc}[HPC]{High Performance Computing}  
  \newacro{moc}[MoC]{Model of Computation}
  \newacroplural{moc}[MoC]{Models of Computation}
  \newacro{ocr}[OCR]{Optical Character Recognition}
  \newacro{qos}[QoS]{Quality of service}
  \newacroplural{qos}[QoS]{Qualities of service}
  \newacro{tpr}[TPR]{True Positive Rate}
  \newacro{mac}[MAC]{Multiply Accumulate}
  \newacro{fc}[FC]{Fully Connected}
  \newacro{simd}[SIMD]{Single Instruction on Multiple Data}
  \newacro{vhdl}[VHDL]{VHSIC Hardware Description Language}
  \newacro{lut}[LUT]{Look-Up Table}
  \newacro{nef}[NEF]{Neighborhood Extraction Factorization}
  \newacro{ne}[NE]{Neighborhood Extraction}
  \newacro{lut}[LUT]{Lookup Table}
  \newacro{hdl}[HDL]{Hardware Description Language}
  \newacro{rtl}[RTL]{Register Transfer Level}
  \newacro{ip}[IP]{Intellectual Property}
  \newacro{dhm}[DHM]{Direct Hardware Mapping}
  \newacro{dag}[DAG]{Direct Acyclic Graph}
  \newacro{dpn}[DPN]{Data-flow Process Network}
  \newacro{sdf}[SDF]{Static Data-Flow}
  \newacro{sdfg}[SDFG]{Synchronous DataFlow Graph}
  \newacroplural{ip}[IP]{Intellectual Properties}
  \newacro{sfp}[SFP]{Static Fixed Point}
  \newacro{dfp}[DFP]{Dynamic Fixed Point}
  \newacro{ttq}[TTQ]{Trained Ternary Quantization}
\usepackage{color}
\definecolor{bluekeywords}{rgb}{0.13,0.13,1}
\definecolor{greencomments}{rgb}{0,0.5,0}
\definecolor{redstrings}{rgb}{0.9,0,0}

\usepackage{array}              % for \newcolumntype macro
\usepackage{multirow}
\newcolumntype{L}{>{$}l<{$}}    % math-mode version of "l" column type
\newcolumntype{C}{>{$}c<{$}}    % math-mode version of "c" column type
\newcommand{\matr}[1]{\bm{#1}}  % Matrix notation
\usepackage{listings}
\lstdefinelanguage{mycpp}{
language=c,
frame = single,
showspaces=false,
showtabs=false,
breaklines=true,
showstringspaces=false,
breakatwhitespace=true,
escapeinside={(*@}{@*)},
commentstyle=\color{greencomments},
keywordstyle=\color{bluekeywords}\bfseries,
stringstyle=\color{redstrings},
basicstyle=\small\fontfamily{fvm}\selectfont
}

\begin{document}
\title{Accelerating CNN inference on FPGAs: A Survey}

\author[1]{Kamel Abdelouahab}
\author[1,2]{Maxime Pelcat}
\author[1]{Jocelyn Sérot}
\author[1]{François Berry}

\affil[1]{Institut Pascal,Clermont Ferrand, France}
\affil[2]{IETR, INSA Rennes, France}
\date{January 2018}
 %%=============================================================
 \maketitle
\begin{abstract}
Convolutional Neural Networks (CNNs) are currently adopted to solve an ever greater number of problems, ranging from speech recognition to image classification and segmentation. The large amount of processing required by CNNs calls for dedicated and tailored hardware support methods. Moreover, CNN workloads have a streaming nature, well suited to reconfigurable hardware architectures such as FPGAs.

The amount and diversity of research on the subject of CNN FPGA acceleration within the last 3 years demonstrates the tremendous industrial and academic interest. This paper presents a state-of-the-art of CNN inference accelerators over FPGAs. The computational workloads, their parallelism and the involved memory accesses are analyzed. At the level of neurons, optimizations of the convolutional and fully connected layers are explained and the performances of the different methods compared. At the network level, approximate computing and datapath optimization methods are covered and state-of-the-art approaches compared. The methods and tools investigated in this survey represent the recent trends in FPGA CNN inference accelerators and will fuel the future advances on efficient hardware deep learning.
\end{abstract}
    \section{Introduction}
The exponential growth of big data during the last decade motivates for innovative methods to extract high semantic information from raw sensor data such as videos, images and speech sequences. Among the proposed methods, \acp{cnn}~\cite{Lecun2015} have become the \textit{de-facto} standard by delivering near-human accuracy in many applications related to machine vision (e.g classification~\cite{Russakovsky2015}, detection~\cite{Girshick2015}, segmentation~\cite{Long2015}) and speech recognition~\cite{Zhang2017d}. 

This performance comes at the price of a large computational cost as \acp{cnn} require up to 38 GOP/s to classify a single frame~\cite{Simonyan2014}. As a result, dedicated hardware is required to accelerate their execution. \acp{gpu}, are the most widely used platform to implement \acp{cnn} as they offer the best performance in terms of pure computational throughput, reaching up 11 TFLOP/s~\cite{Nurvitadhi2017}. Nevertheless, in terms of power consumption, \ac{fpga} solutions are known to be more energy efficient (vs \acp{gpu}). As a result, numerous FPGA-Based CNN accelerators have been proposed, targeting both \ac{hpc} data-centers~\cite{Ovtcharov2015} and embedded applications~\cite{Qiu2016}.

While GPU implementations have demonstrated state-of-the-art computational performance, CNN acceleration is shortly moving towards FPGAs for two reasons. First, recent improvements in FPGA technology put FPGA performance within striking distance to GPUs with a reported performance of 9.2 TFLOP/s for the latter~\cite{IntelFPGA2017a}. 
Second, recent trends in CNN development increase the sparsity of CNNs and use extreme compact data types. These trends favorize FPGA devices which are designed to handle irregular parallelism and custom data types. As a result, next generation CNN accelerators are expected to deliver up to x5.4 better computational throughput than GPUs.~\cite{Nurvitadhi2017}. 

As an inflection point in the development of CNN accelerators might be near, we conduct a survey on FPGA-Based CNN accelerators. While a similar survey can be found in~\cite{Lacey2016}, we focus in this paper on the recent techniques that were not covered in the previous works. Moreover, a recent review of efficient processing techniques for deep learning is proposed in~\cite{Sze2017}, but focuses on \ac{asic} accelerators for CNNs while our work is mainly related to FPGA-based implementations. 

The rest of the paper is organized as follows, section~\ref{sec:Background} recalls the main features of CNNs, focusing on computations and workload issues. Section~\ref{sec:OptAlgo} studies the computational transforms exploited to accelerate CNNs on FPGAs. Section~\ref{sec:OptDP} reviews the contributions that attempt to optimize the data-path of FPGA-Based CNN accelerators. Section~\ref{sec:OptAC} shows how approximate computing is a key in the acceleration of CNNs on FPGAs and overviews the main contributions implementing these techniques. Finally, section 6 concludes the paper.%, section~\ref{sec:OptGen} demonstrates how \ac{hls} is employed increases the productivity to generate CNN accelerators and compares the performance \ac{hls} based and low level \ac{hdl}-base CNN accelerators on FPGAs.

    %\newpage
\section{Background on CNNs}
\label{sec:Background}

This section overviews the main features of \acp{cnn} and focuses on the computations and parallelism patterns involved during their inference.

\subsection{General Overview:}
\label{sec:Overview}
\acp{cnn} are feed-forward, deep, sparsely connected neural networks that implement weight sharing. A typical \ac{cnn} structure consists of a pipeline of layers. Each layer inputs a set of data, known as a \acf{fm}, and produces a new set of \acp{fm} with \emph{higher-level semantics}.

% Describe the CNN topolgy
% \begin{figure}[ht]
% \centering
% \includegraphics[width=.7\textwidth]{img/cnnStructure.eps}
% \caption{A typical CNN Structure with 3 convolutional layers, 2 \emph{pool} layers and a fully connected stage}
% \label{cnnStructure}
% \end{figure}

\subsection{Inference \textit{vs} Training:}
\label{sec:TrainInfer}

As typical \ac{ml} algorithms, \acp{cnn} are deployed in two phases. First, the \emph{training} stage works on a known set of annotated data samples to create a model with a \textit{modeling} power (i.e. which semantics extrapolates to natural data outside the training set). This phase implements the \textit{back-propagation} algorithm~\cite{Lecun1998} which iteratively updates \ac{cnn} parameters such as convolution weights to improve the predictive power of the model. CNN Models can also be \textit{fine-tuned}. When \textit{fine-tuning} a model, weights of a previously-trained network are used to initialize the parameters of a new training. These weights are then adjusted for a new constrain, such as a different dataset or a reduced precision. 

The second phase, known as \emph{inference}, uses the learned model to classify new data samples (i.e inputs that were not previously seen by the model). In a typical setup, CNNs are trained/fine-tuned only once, on large GPU/FPGA clusters. By contrast, the inference is implemented each time a new data sample has to be classified. As a consequence, the literature mostly focuses on accelerating the inference phase. As a result, this paper overviews the main methods employed to accelerate the inference\footnote{The computational transforms discussed in sections~\ref{sec:OptAlgo} and approximate computing techniques detailed in section~\ref{sec:OptAC} can both be employed during the training and the inference.}. Moreover, since most of the CNN accelerators benchmark their performance on models trained for image classification, we focus on this paper on this application. Nonetheless, the methods studied in this survey can be employed to accelerate CNNs for other applications such object detection, image segmentation and speech recognition.

% In this survey, the computational transforms detailed in section~\ref{sec:OptAlgo}, Since the training phase is only implemented once such FFT, can be employed in both 

% From a computational point of view, the learning phase requires several orders of magnitude more computation than inference. In addition, it requires high precision arithmetic to support the optimization algorithms and gradient computation employed in back-propagation \textcolor{red}{[need for references]}. As a result, the training of a \ac{cnn} is typically performed off-line on large clusters of GPUs/FPGAs. The Acceleration of the training phase is kept beyond the scope of this survey, \Rework{\ac{fpga} implementations of back-propagation can be found in~\cite{Eldredge1994,Ortega2016,Ko2017}}.

\subsection{Inference of \acp{cnn}}
\label{sec:Inference}
\ac{cnn} inference refers to the \emph{feed-forward} propagation of $B$ input images across $L$ layers. This section details the computations involved in the major types of these layers. A common practice is to manipulate layer parameters and \acp{fm} using tensors. The tensors and variables used in this work are listed in table~\ref{Notation}.

\begin{table}[ht]
\caption{Tensors Involved in the inference of a given layer $\ell$ with their dimensions}
\centering
\label{Notation}
\begin{tabular}{|C|c|C|}
\hline
\displaystyle \matr{X} 		& Input \acp{fm} 	& B\times C\times H\times W\\ \hline
\displaystyle \matr{Y}		& Output \acp{fm}	& B\times N\times V\times U\\ \hline
\displaystyle \matr{\Theta}	& Learned Filters	& N\times C\times J\times K \\ \hline
\displaystyle \beta			& Learned biases	& N\\ \hline
\end{tabular}
\begin{tabular}{|C|c|}
\hline
\displaystyle B  			& Batch size (Number of input frames)\\ \hline
%\displaystyle L  			& Number of Layers (Depth of CNN)	\\ \hline
\displaystyle W / H / C		& Width / Height / Depth of Input \acp{fm} 	\\ \hline
\displaystyle U / V / N		& Width / Height / Depth of Output \acp{fm}	\\ \hline
% \displaystyle C 			& Depth of input \acp{fm} 			\\ \hline
% \displaystyle N 			& Depth of output \acp{fm}			\\ \hline
\displaystyle K / J			& Horizontal / Vertical Kernel size	\\ \hline
\end{tabular}
\end{table}

\subsubsection{Convolution layers:\\} 
\begin{figure}[ht]
\centering
\includegraphics[width=.8\textwidth]{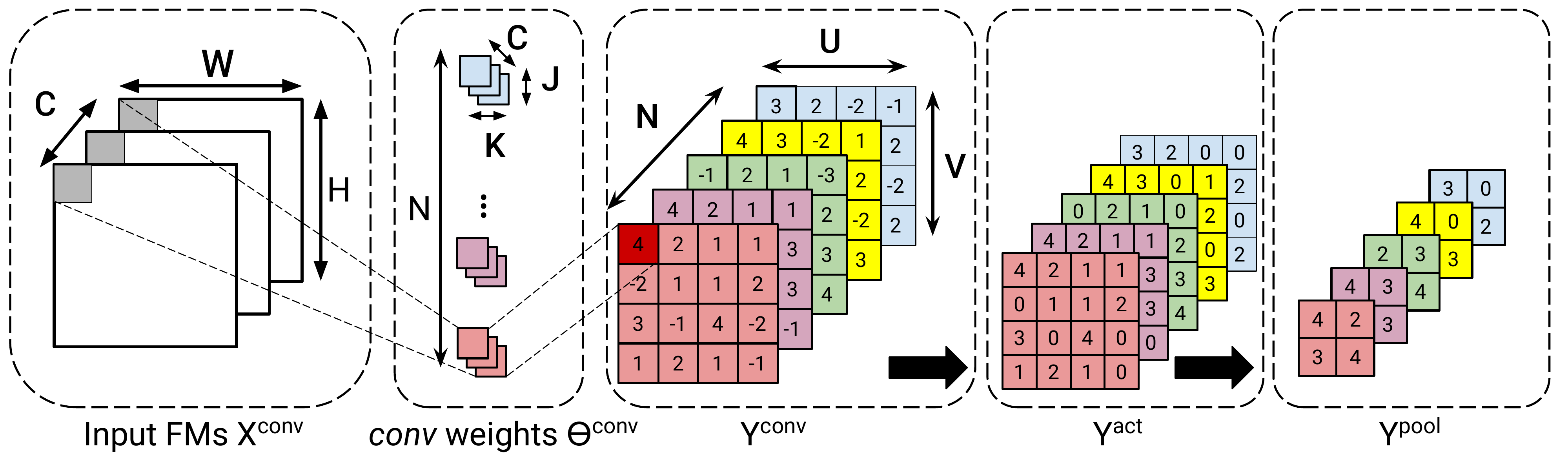}
%\subfloat[]{\label{convImg} \resizebox{0.65\textwidth}{!}{\input{tikz/conv.tex}}}
%\subfloat[]{\label{poolImg} \includegraphics[width=0.35\textwidth]{img/maxpool.jpeg}}
\caption{Feed forward propagation in \textit{conv}, \textit{act} and \textit{pool} layers (Batch size $B$=1, bias $\beta$ omitted)}
\label{img:FeedForward}
\end{figure}

A convolution layer (\textit{conv}) carries out the feature extraction process by applying --as illustrated in figure~\ref{img:FeedForward}-- a set of 3D-convolution filters ${\matr{\Theta}^{\text{conv}}}$ to a set of $B$ input volumes $\matr{X}^{\text{conv}}$. Each input volume has a depth $C$ and can be a color image (in the case of the first \textit{conv} layer), or an output generated by previous layers in the network. Applying a 3D-filter to 3D-input results in a 2D \emph{\acf{fm}} and, each \textit{conv} layer outputs a set of $N$ two-dimensional features maps.
In some \ac{cnn} models, a learned offset $\beta^{\text{conv}}$ --called a \emph{bias}-- is added to the 3D-conv results, but this practice is discarded in recent models~\cite{Simonyan2014}. The computations involved in feed-forward propagation of \textit{conv} layers are detailed in equation~\ref{convLayer}.

\begin{align}
\label{convLayer}
\forall & \left\{ b, n, u, v \right\} \in \left[ 1,B \right] \times \left[ 1,N \right] \times \left[ 1,V \right] \times \left[ 1,U \right] \nonumber \\
		& \matr{Y}^{\text{conv}}[b,n,v,u] = \beta^{\text{conv}}[n] + \sum_{c=1}^{C^{}} \sum_{j=1}^{J}  \sum_{k=1}^{K} \matr{X}^{\text{conv}}[b,c,v+j,u+k] . \matr{\Theta}^{\text{conv}}[n,c,j,k]
\end{align}

%\hl{- Sparse connections}\\

\subsubsection{Activation Layers:\\} 
Each \textit{conv} layer of a \ac{cnn} is usually followed by an activation layer that applies a \emph{non-linear} function to all the values of \acp{fm}. Early \acp{cnn} were trained with TanH or Sigmoid functions but recent models employ the \ac{relu} function that grants faster training times and less computational complexity, as highlighted in~\cite{Krizhevsky2012a}.
\begin{align}
\label{actLayer}
\forall & \left\{ b, n, u, v \right\} \in \left[ 1,B \right] \times \left[ 1,N \right] \times \left[ 1,V \right] \times \left[ 1,U \right] \nonumber \\
		& \matr{Y}\textsuperscript{act} [b,n,h,w] = \text{act} \Big(  \matr{X}\textsuperscript{act} [b,n,h,w] \Big) \qquad \ | \qquad \text{ act := TanH, Sigmoid, ReLU ...}
\end{align}
\subsubsection{Pooling layers:\\}
The convolutional and activation parts of a \ac{cnn} are directly inspired by the cells of visual cortex in neuroscience~\cite{Hubel1962}. This is also the case of \emph{pooling} layers, which are periodically inserted in-between successive \textit{conv} layers. As shown in equation~\ref{poolLayer}, \emph{pooling} sub-samples each channel of the input \acp{fm} by selecting the \emph{average}, or, more commonly, the \emph{maximum} of a given neighborhood $K$. As a results, the dimensionality of a \acp{fm} is reduced, as illustrated in figure~\ref{img:FeedForward}
\begin{align}
\label{poolLayer}
\forall & \left\{ b, n, u, v \right\} \in \left[ 1,B \right] \times \left[ 1,N \right] \times \left[ 1,V \right] \times \left[ 1,U \right] \nonumber \\
    	& \matr{Y}\textsuperscript{pool} [b,n,v,u] = \max_{p,q \in [1:K]}{\left(\matr{X}\textsuperscript{pool} [b,n,v+p,u+q]\right)}
\end{align}

\subsubsection{Fully Connected Layers:\\}
When deployed for classification tasks, the \acp{cnn} pipeline is often terminated by \acf{fc} layers. These layers can be seen as \textit{conv} layers with no weight sharing (i.e $W=K$ and $H=J$). Moreover, in a same way as \textit{conv} layers, a non-linear function is applied to the outputs of \textit{FC} Layers.
\begin{align}
\label{fcLayer}
\forall & \left\{ b, n \right\} \in \left[ 1,B \right] \times \left[ 1,N \right] \nonumber \\
		&\matr{Y}^{\text{fc}}[b,n] = \beta^{\text{fc}}[n] + \sum_{c=1}^{C} \sum_{h=1}^{H}  \sum_{w=1}^{W} \matr{X}^{\text{fc}}[b,c,h,w] . \matr{\Theta}^{\text{fc}}[n,c,h,w]
\end{align}

\subsubsection{Batch-Normalization Layers:\\}
\label{sec:BatchNorm}
Batch-Normalization is introduced in~\cite{Ioffe2015} to speed up training by linearly shifting and scaling the distribution of a given batch of inputs $B$ to have zero mean and unit variance. These layers find also there interest when implementing \ac{bnn} (cf section~\ref{sec:bnn}) by reducing the quantization error compared to an arbitrary input distribution, as highlighted in~\cite{Courbariaux2016}. Equation~\ref{eq:BatchNorm} details the processing of \textit{batch norm} layers, where $\mu$ and $\sigma$ are statistic collected during the training, $\alpha$ , $\epsilon$ and $\gamma$ parameters are training hyper-parameters.

\begin{align}
\label{eq:BatchNorm}
\forall & \left\{ b, n, u, v \right\} \in \left[ 1,B \right] \times \left[ 1,N \right] \times \left[ 1,V \right] \times \left[ 1,U \right] \nonumber \\
		& \matr{Y}^{\text{BN}}[b,n,v,u] = \frac{\matr{X}^{\text{BN}}[b,n,u,v] - \mu}{\sqrt{\sigma^2+\epsilon}}\gamma + \alpha 
\end{align}

\subsection{Workload of a \acp{cnn} inference}
\label{sec:Workload}

\begin{table}[ht]
\caption[Popular CNN Models]{Popular CNN models with their computational workload. Accuracy measured on single-crops of ImageNet test-set.}
\label{tab:PopularCNNs}
\resizebox{\textwidth}{!}{
\begin{tabular}{|c|c|c|c|c|c|c|c|}
\hline
Model                 & AlexNet~\cite{Krizhevsky2012a}  & GoogleNet~\cite{Szegedy2015} & VGG16~\cite{Simonyan2014}   & VGG19~\cite{Simonyan2014}   & ResNet50~\cite{He2016b} & ResNet101~\cite{He2016b} & ResNet-152~\cite{He2016b} \\ \hline
Top1 err              & 42.9 \%  & 31.3 \%   & 28.1 \% & 27.3 \% & 24.7\%   & 23.6\% \% & 23.0\%     \\ \hline
Top5 err              & 19.80 \% & 10.07 \%  & 9.90 \% & 9.00 \% & 7.8 \%   & 7.1 \%    & 6.7 \%     \\ \hline \hline
conv layers           & 5        & 57        & 13      & 16      & 53       & 104       & 155        \\ \hline
conv workload (MACs)  & 666 M    & 1.58 G    & 15.3 G  & 19.5 G  & 3.86 G   & 7.57 G    & 11.3 G     \\ \hline
conv parameters       & 2.33 M   & 5.97 M    & 14.7 M  & 20 M    & 23.5 M   & 42.4 M    & 58 M       \\ \hline \hline
Activation layers     & \multicolumn{7}{c|}{\ac{relu}}                                              \\ \hline
pool layers           & 3        & 14        & 5       & 5       & 2        & 2         & 2          \\ \hline \hline
FC layers             & 3        & 1         & 3       & 3       & 1        & 1         & 1          \\ \hline
FC workload (MACs)    & 58.6 M   & 1.02 M    & 124 M   & 124 M   & 2.05 M   & 2.05 M    & 2.05 M     \\ \hline
FC parametrs          & 58.6 M   & 1.02 M    & 124 M   & 124 M   & 2.05 M   & 2.05 M    & 2.05 M     \\ \hline \hline
Total workload (MACs) & 724 M    & 1.58 G    & 15.5 G  & 19.6 G  & 3.86 G   & 7.57 G    & 11.3 G     \\ \hline
Total parameters      & 61 M     & 6.99 M    & 138 M   & 144 M   & 25.5 M   & 44.4 M    & 60 M       \\ \hline
\end{tabular}
}
\end{table}

The accuracy of \ac{cnn} models have been increasing since their breakthrough in 2012~\cite{Krizhevsky2012a}. However, this accuracy comes at the price of a high computational cost. The main challenge that faces \ac{cnn} developers is to improve classification accuracy while maintaining a tolerable computational workload. As shown in table~\ref{tab:PopularCNNs}, this challenge was successfully addressed by Inception~\cite{Szegedy2015} and ResNet models~\cite{He2016b}, with their use of bottleneck $1\times1$ convolutions that reduce both model size and computations while increasing depth and accuracy.

\subsubsection{Computational Workload:\\ }
\label{Workload:comp}
\label{Workload:ParamSize}
The computational workload of a \ac{cnn} inference is the result of an intensive use of the \acf{mac} operation. Most of these \acp{mac} occur on the convolutional parts of the network, as shown in tab~\ref{tab:PopularCNNs}. As a consequence, \textit{conv} layers are responsible, in a typical implementation, of more than 90\% of execution time during the inference~\cite{Cong2014}. Conversely to computations, and as shown in tab~\ref{tab:PopularCNNs}, most of the \ac{cnn} weights are included on the \ac{fc}-layers. Due to this unbalanced computation to memory ratio, \acp{cnn} accelerators follow different strategies when implementing the convolutional and fully connected parts of inference. 

% The number of \acp{mac} involved in a CNN inference ($\mathscr{C}$) is computed as follows:
% \begin{align}
% \mathscr{C} = & \sum_{\ell=0}^{L^{\text{conv}}} \mathscr{C}^{\text{conv}}_{\ell} + \sum_{\ell=0}^{L^{\text{fc}}} \mathscr{C}^{\text{fc}} \hspace{0.7cm} \text{(MACs)}\label{MACS:total}\\
% & \mathscr{C}_{\ell}^{\text{conv}} = V_{\ell} U_{\ell} K_{\ell}^{2} N_{\ell} C_{\ell} \label{MACS:conv}\\
% & \mathscr{C}_{\ell}^{\text{fc}}   = H_{\ell} W_{\ell} N_{\ell} C_{\ell} \label{MACS:fc}
% \end{align}
\subsubsection{Parallelism in CNNs:\\}
\label{Workload:parallelism}
Because of the high number of required computations, inferring \acp{cnn} with real-time constraints is a challenge, especially on low-energy embedded devices. A solution to this challenge is to take advantage of the extensive concurrency exhibited by \acp{cnn}. These sources can be formalized as:
\begin{itemize}
\item \textbf{Batch Parallelism:} \ac{cnn} implementations can simultaneously classify multiple frames grouped as a \emph{batch} $B$ in order to reuses the filters in each layer and minimize the external memory accesses. As a result, the inference benefits from a significant acceleration when implementing batch processing.
\item \textbf{Inter-layer Parallelism:} \acp{cnn} have a feed-forward hierarchical structure consisting of a succession of data-dependent layers. These layers can be executed in a pipelined fashion by launching layer $(\ell)$ before ending the execution of layer $(\ell-1)$.
\end{itemize}

Moreover, the computation of each \textit{conv} layer, described in eq~\ref{convLayer}, exhibits four sources of concurrency that are detailed above.

\begin{itemize}
\item \textbf{Inter-FM Parallelism:} Each output \ac{fm} plane of a \textit{conv} layer can be processed separately from the others. This means that $P_N$ elements of $\matr{Y}^{\text{conv}}$ can be computed in parallel ($0<P_N<N$).
\item \textbf{Intra-FM Parallelism:} Multiple pixels of a single output \ac{fm} plane can be processed concurrently by evaluating $P_V \times T_U$ Values of $\matr{Y}^{\text{conv}}[n]$ ($0 < P_V \times P_U < V \times U$)
\item \textbf{Inter-convolution Parallelism:} 3D-convolutions occurring in \textit{conv} layers can be expressed as a sum of 2D convolutions as shown in equation~\ref{convLayer2D}. These 2D convolutions can be evaluated simultaneously by computing concurrently $P_C$ elements of eq~\ref{convLayer2D} ($0<P_C<C$).
\item \textbf{Intra-convolution Parallelism:} The 2D-convolutions involved in the processing of \textit{conv} layers can be implemented in a pipelined fashion as in~\cite{Shoup1994}. In this case $P_J \times P_K$ multiplications are implemented concurrently ($0<P_J \times P_K <J \times K$).
\end{itemize}

\begin{align}
    \label{convLayer2D}
    \forall & \left\{ b, n \right\} \in \left[ 1,B \right] \times \left[ 1,N \right] \nonumber \\
    & \matr{Y}^{\text{conv}}[n] = {b [n]} + \sum_{c=1}^{C^{}} \text{\textbf{conv2D}} \Big( {\matr{X}^{\text{conv}}[c]}, {\matr{\Theta}^{\text{conv}}[n,c]} \Big)
\end{align}

%\hl{- Add Figures to explain parallelism?}\\

%\subsubsection{Parameters and model-size of CNNs:\\}

%The number of CNN parameters ($\mathscr{M}$) can be computed according to equation~\ref{Mem:total}.
% \begin{align}
% \mathscr{M} = & \sum_{\ell=0}^{L^{\text{conv}}} \mathscr{M}_{\ell}^{\text{conv}} + \sum_{\ell=0}^{L^{\text{fc}}} \mathscr{M}_{\ell}^{\text{fc}}\label{Mem:total}\\
% & \mathscr{M}_{\ell}^{\text{conv}} = K_{\ell}^{2} N_{\ell} C_{\ell}\label{Mem:conv} \\
% & \mathscr{M}_{\ell}^{\text{fc}}   = H_{\ell} W_{\ell} N_{\ell} C_{\ell}\label{Mem:fc}
% \end{align}

% \begin{align}
% \mathscr{A} = & \sum_{\ell=0}^{L^{\text{conv}}} \mathscr{A}_{\ell}^{\text{conv}} + \sum_{\ell=0}^{L^{\text{fc}}} \mathscr{A}_{\ell}^{\text{fc}}\label{Access:total}\\
% & \mathscr{A}_{\ell}^{\text{conv}} = N_{\ell} V_{\ell} U_{\ell} (2 K_{\ell}^{2} C_{\ell} + 1)\label{Access:conv} \\
% & \mathscr{A}_{\ell}^{\text{fc}}   = N_{\ell} (2 H_{\ell} W_{\ell} C_{\ell} + 1)\label{Access:fc}
% \end{align}

\subsubsection{Memory Accesses in CNNs:\\}
\label{Workload:Memory}
The CNN inference shows large vectorization opportunities that are exploited by allocating multiple computational resources to accelerate the processing. However, this method may be inefficient if no caching strategy is implemented.

In fact, memory bandwidth is often the bottleneck when processing \acp{cnn}. For the \textit{FC} parts, execution can be memory-bounded because of the high number of weights that these layers contain, and consequently, the high number of memory reads engendered.  For the \textit{conv} parts, the high number of \ac{mac} operations results in a high amount of memory accesses because each \ac{mac} requires at least 2 memory reads and 1 memory write to be performed\footnote{This is the best case scenario of a fully pipelined MAC where intermediate results don't need to be loaded.}. If all these accesses are towards external memory (for instance, \ac{dram}), throughput and energy consumption will be highly impacted since a \ac{dram} access engenders significantly more of latency and energy consumption than the computation it self~\cite{Horowitz2014}

%\begin{equation}
%\mathscr{M}_{\ell} = 3 \mathscr{C}_{\ell}  \label{MemAccess} \\
%\end{equation}

The number of these DRAM accesses, and thus latency and energy consumption, can be reduced by implementing a memory caching hierarchy using on-chip memories. As discussed in section~\ref{sec:OptDP}, Hardware accelerators for \acp{cnn} usually employ two levels of caches. The first level is implemented by means of large on-chip buffers while the second level involves local register files implemented at the nearest of the computational capabilities. The latency and energy consumption that result from memory access toward these 2 cache levels is several order of magnitude less then external memory access, as pointed-out in~\cite{Sze2017}.\\

\subsubsection{Hardware, libraries and frameworks:\\}
% \begin{itemize}
% \item \Plan{In order to catch the parallelism of CNNs, Dedicated hardware, such many-core CPUs and GPUs are needed }
% \item \Plan{Specialized libraries are developed to develop CNNs on dedicated hardware while providing abstraction: CudNN on nVidia GPUs, BigDL on Intel CPUs, DeepCL on heterogeneous hardware through OpenCL standard}
% \item \Plan{Frameworks are built-upon these libraries to improve productivity of crafting, training and deploying CNNs. The main frameworks are Caffe\cite{Jia2014} and TensorFlow~\cite{Abadi2016}}
% \end{itemize}
In order to catch the parallelism of CNNs, dedicated hardware accelerators are developed. Most of them are based on \ac{gpu}, which that are known to perform well on regular parallelism patterns thanks to a \ac{simd} and \ac{simt} execution models, a dense collection of floating-point computing elements that peaks at 12 TFLOPs, and high capacity/bandwidth on/off-chip memories~\cite{Nvidia2015}.
To support these hardware accelerators, specialized libraries for deep learning are developed to provide the necessary programming abstraction, such as CudNN on Nvidia \acp{gpu}~\cite{cudnn14} and DeepCL on heterogeneous hardware through OpenCL standard\cite{Perkins2017}. Built-upon these libraries, dedicated frameworks for deep learning are proposed to improve productivity of conceiving, training and deploying CNNs, such as Caffe\cite{Jia2014} and TensorFlow~\cite{Abadi2016}.\\

%\Plan{Transition avec les FPGA : voir commentaires}\\
Beside \ac{gpu} implementations, numerous \ac{fpga} accelerators for \acp{cnn} have been proposed. \acp{fpga} are fine-grain programmable devices that can catch the CNN parallelism patterns with no memory bottleneck, tanks to
\begin{enumerate}
\item A High density of hard-wired \ac{dsp} blocs that are able to achieve up to 20 (8 TFLOPs) TMACs~\cite{IntelFPGA2017a}.
\item A collection of \textit{In-situ} on-chip memories, located next to DSPs, that can be exploited to significantly reduce the number of external memory accesses.
\end{enumerate}

% \subsubsection{The case of FPGAs:\\}
% \Plan{Transition: Fine grain devices such FPGAs can take advantage of this parallelism with no memory bottleneck because}\\
% \begin{itemize}
% 	\item \Plan{Lecher les bottes de l'editeur : Citer \cite{Putnam2014} }
% 	\item \Plan{FPGA technology is evolving}
% 	\subitem \Plan{- FPGAs embed \textit{in-situ} on-chip memories to reduce external memory access}
% 	\subitem \Plan{- FPGAs embed high density hard-wired \acp{dsp} blocs that are able to support up to 20 (8 TFLOPs) TMACs~\cite{IntelFPGA2017a}}
% 	\item \Plan{Reconfigurability vs ASICs,Energy efficiency vs CPUs/GPUs}
% 	\item \Plan{FPGAs have been known to have better energy efficiency compared to GPUs, but the latter outperforms the former in terms of computational throughput and "productivity" }
% 	\item \Plan{As pointed-out in~\cite{Nurvitadhi2016b,Nurvitadhi2017},this is changing because}
% 	\subitem \Plan{- New CNN models exploits network sparsity\cite{Venkatesh2017,Kim2017,Sironi2015,Han2015b,Iandola2016,Hong2015} = Irregular parallelism = where FPGAs shine}
% 	\subitem \Plan{- CNN development exploits extreme compact data-types that are handled in a more efficient way on FPGAs \cite{Li2016,Guo2017,Courbariaux2014,Kim2017,Moons2016a,Darryl2016,Rastegari2017,Zhou2016b,Suyog2015,Courbariaux2015}}
% 	\subitem \Plan {- HLS tools and OpenCL are evolving rapidly which makes productivity of FPGA programming comparable to GPUs }
% 	\subitem \Plan{An inflection point is near !}
% \end{itemize}
When porting a \ac{cnn} to an \ac{fpga} device, the problem boils down to finding an efficient mapping between the computational model of the former and the execution model supported by the latter. In the the following sections, the main strategies explored by the literature to address this mapping problem are reviewed. In particular, we show that current \ac{fpga}-based accelerators for \acp{cnn} rely on one (or a combination) of three main optimizations to efficiently infer \acp{cnn}.

%%\begin{landscape}
\begin{figure}[ht]
\begin{adjustbox}{width=.9\linewidth}
	\begin{forest}
for tree={%
    l sep=1cm,
    s sep=0.1cm,
    minimum height=0.8cm,
    minimum width=1cm,
    rounded corners=2pt,
    align=center,
    draw
    }
    [FPGA Acceleration of CNNs
		[Algorithmic Optimization
           [GEMM\\~\cite{Cong2014,Suda2016,Nurvitadhi2017,Chellapilla2006}]
			[Winograd\\~\cite{Aydonat2017,DiCecco2016}]
			[FFT\\~\cite{Zhang2017,Ko2017}]
		]
		[Datapath Optimization
			[SDF / DPN \\ \cite{Venieris2016,Sharma2016}\\ \cite{Li2016a,Natale2017,Abdelouahab2017}]
			[DSE/Roofline \\ \cite{Zhang2015,Motamedi2016}\\ \cite{Suda2016,Meloni2016}\\ \cite{Motamedi2017,Wei2017}]
			[misc.]
		]
		[CNN model Optimization
        	[Sparsity
            	[Pruning \\ \cite{Molchanov17,Fujii2017}]
            	[SVD \\ \cite{Qiu2016}]
        	]
        	[Quantization
            	[Linear \\ \cite{Suyog2015, Zhou2016b} \\ \cite{Courbariaux2014,Qiu2016,Gysel2016}]
            	[Binary \\ \cite{Courbariaux2015,Rastegari2017} \\ \cite{Umuroglu2017,Andri2016,Zhao2015}]
        	]
        	[Stochastic\\~\cite{Sim2017b,Lee2017}\\~\cite{Vogel,Kim2016b,Ren2016,Ardakani2015}]
      	]
      	[Hardware Generation
      		[HLS Based
      			[OpenCL\\ \cite{Suda2016} \\ \cite{Aydonat2017,Zhang2017a}]
      			[Vivado HLS\\ \cite{Zhang2015,Umuroglu2017}\\ \cite{Venieris2016} \\~\cite{Tapiador2017}]
      		]
      		[DSL Based]
      		[RTL \\ \cite{Motamedi2017}]
      	]
      ]
\end{forest}
\end{adjustbox}
\caption{Main Approaches to Accelerate \ac{cnn} inference on \acp{fpga}}
\end{figure}
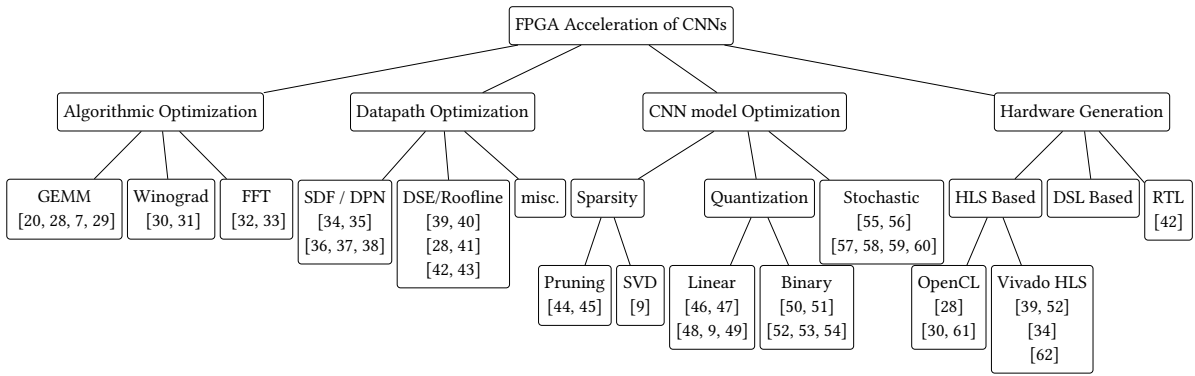
%%\end{landscape}

    \section{Algorithmic Optimizations for FPGA-Based CNN Acceleration}
\label{sec:OptAlgo}
%Inescapably, when evaluating hardware platforms to accelerate a domain specific application, the trade-off between flexibility, performance and power consumption is always considered. On one end of the spectrum, \acp{gpp} provide a high degree of programmability while performing relatively inefficiently (in terms of performance/watt ratio). On the other end of the spectrum, \acp{asic} delivers better performance per watt while being less flexible and more expensive to produce. \acp{fpga} stay between the two and delivers a good compromise between the three metrics. This is especially true for \ac{cnn} inference acceleration; While \acp{fpga} have not been known for offering top performance when compared to \acp{gpp} and \acp{gpu}, they have provided superior energy efficiency (vs \acp{gpu}) and better flexibility (vs \acp{asic}).\\

% \begin{itemize}
% \item \Plan{Computational transforms that reduces the number of multiplications. Increases throughput but not so energy efficient}
% \item \Plan{Even if it is widely used by CPU/GPU community, \ac{fpga} can take advantage of these techniques}
% \end{itemize}
%\Plan{Change the Algorithmic name}\\
In order to accelerate the execution of \textit{conv} and \ac{fc} layers, computational transforms are employed on the \acp{fm} and kernels in order to vectorize the implementations and reduce the number of arithmetic operations occurring during inference. These computational transforms are mainly deployed in CPUs and \ac{gpu} and are implemented by means of variety of software libraries such OpenBlas CPUs and cuBLAS for \acp{gpu}. Beside this, various implementations make use of such transforms to map \acp{cnn} on \acp{fpga}.

\subsection{GEMM Transformation}
\label{sec:GEMM}
In \acp{cpu} and \acp{gpu}, a common way to process \acp{cnn} is to map \textit{conv} and FC layers as \acp{gemm}. The OpenCL standard generalizes this approach to \acp{fpga}-based implementations~\cite{Bottleson2016,IntelFPGA2016}.\\

\begin{figure}[ht]
	\centering
	\subfloat[]{\label{fig:GEMM-FC}\includegraphics[width=0.4\textwidth]{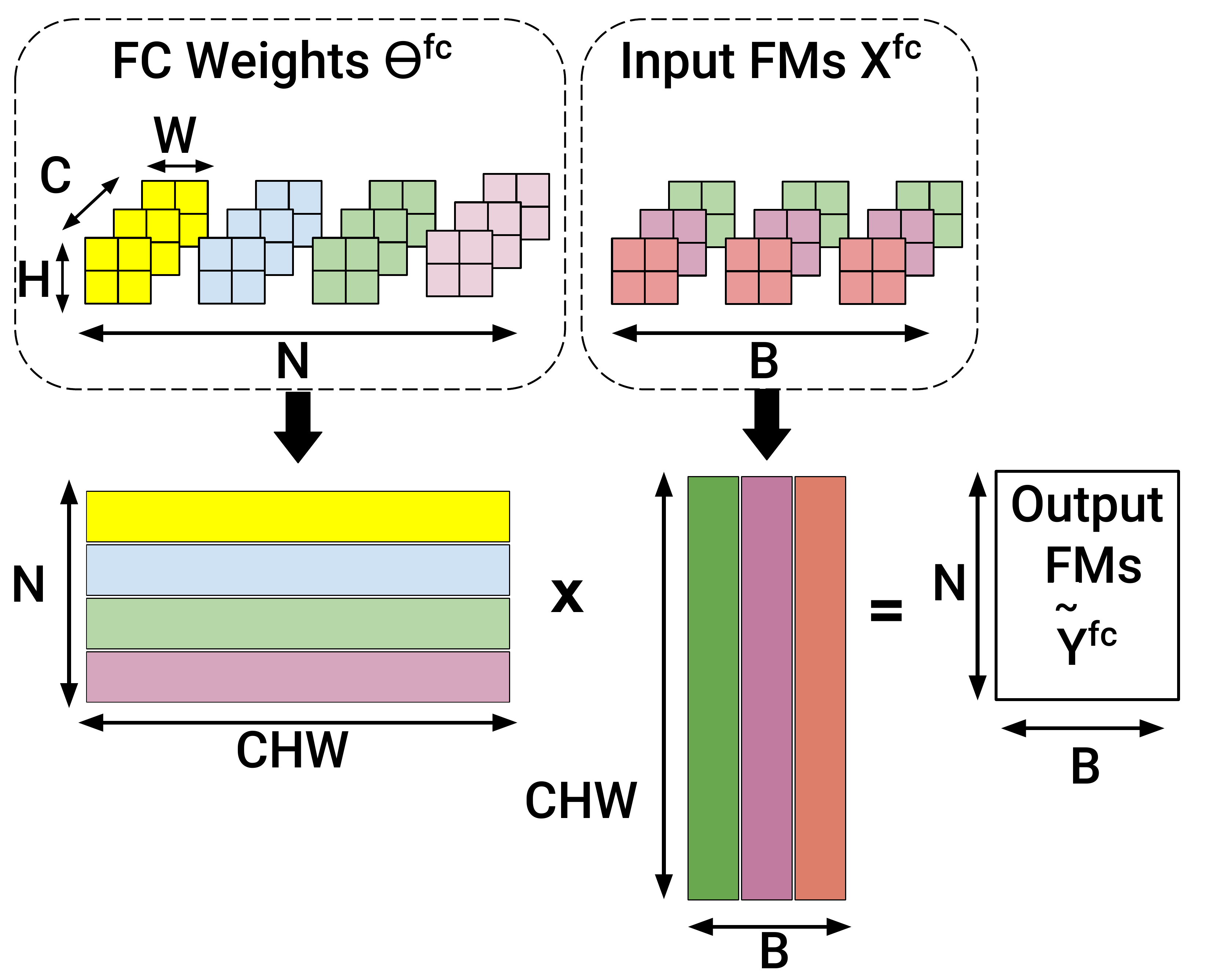}}
    \hfill
	\subfloat[]{\label{fig:GEMM-Conv}\includegraphics[width=0.4\textwidth]{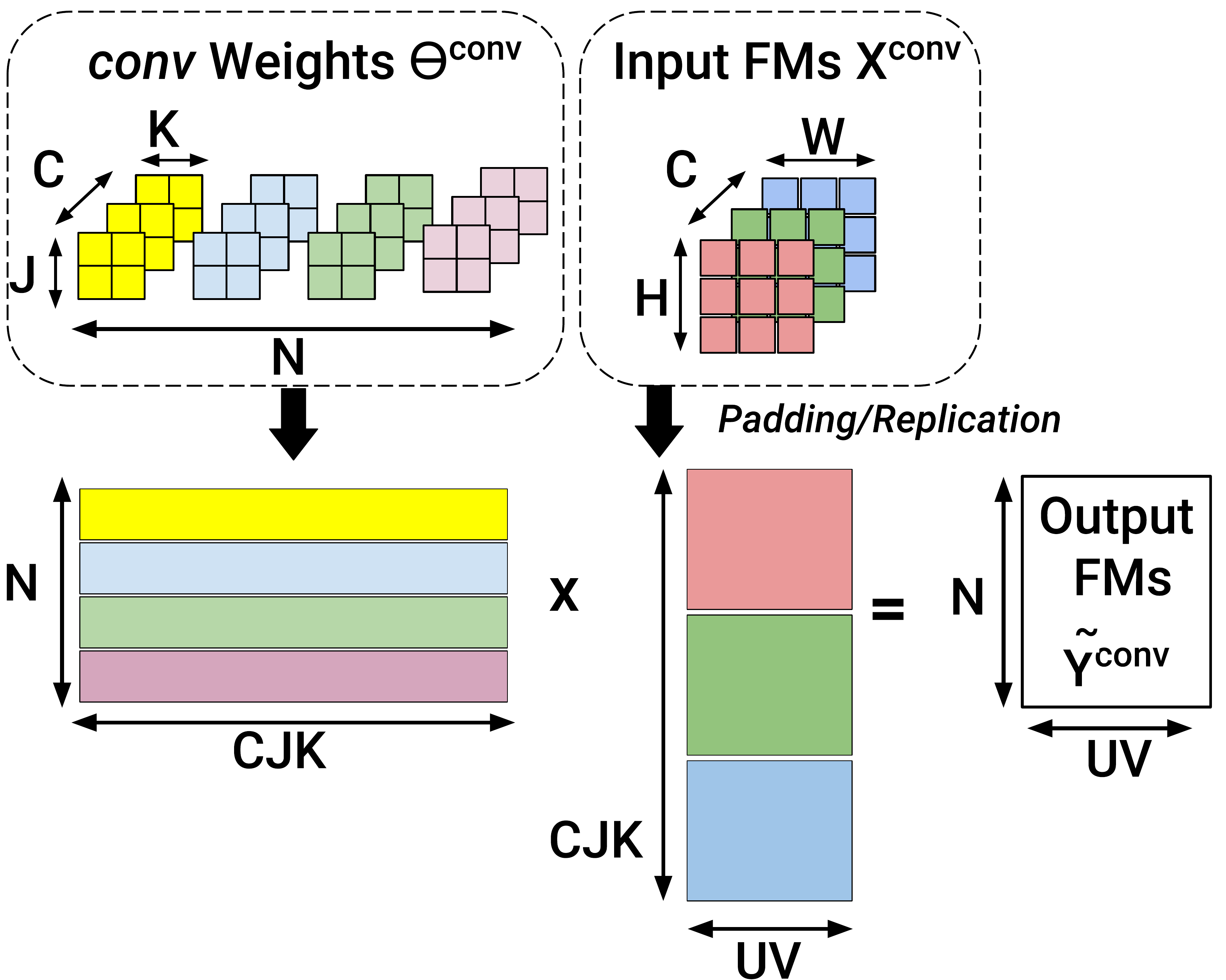}}
	\caption{GEMM Based processing of: a- FC layers, b- conv layers.}
	\label{fig:GEMM}
\end{figure}

%\subsubsection{GEMM Implementation of FC layers:}
\label{sec:GEMM-FC}
For \ac{fc} layers, in which the processing boils down to a matrix-vector multiplication problem, the \ac{gemm}-based implementations find its interest when processing a \emph{batch} of \acp{fm}. In this case, the batch is concatenated onto a $CHW \times B$ matrix, as shown in fig~\ref{fig:GEMM-FC}.

As mentioned in section~\ref{Workload:ParamSize}, most of the weights of CNNs are employed in the FC parts. Instead of loading these weights multiple times to classify multiple inputs, feature maps of FC layers are \textit{batched} in a way that FC weights are loaded only one time per batch. This vectorization is employed in~\cite{Zhang2016,Zhang2016a,Aydonat2017} to increase the computational throughput in FC layers while maintaining a constant memory bandwidth utilization. Moreover, the efficiency of this method increases as the sparsity of $\matr{\Theta}^{\text{fc}}$ grows (cf. sec~\ref{sec:sparcity}).

%\subsubsection{GEMM Implementation of conv layers:}
3D Convolutions can also be mapped as \acp{gemm} using, for instance, the computational transform introduced in~\cite{Chellapilla2006}. Suda et \textit{al.}~\cite{Suda2016} and more recently, Zhang et \textit{al.}~\cite{Zhang2017a} leverage on this \acp{gemm} transcription of 3D convolution to derive OpenCL-based \ac{fpga} Accelerators for \acp{cnn}.
In these works, a transformation flattens all the filters of a given \textit{conv} layer onto an $N \times CKJ$ matrix ${\tilde{\Theta}}$ and re-arranges input \acp{fm} onto a $CKJ \times UV$ matrix ${\tilde{X}}$. The output \acp{fm}, ${\tilde{Y}}$, is the result of the multiplication of the two former matrices, as illustrated in Fig~\ref{fig:GEMM-Conv}.
The mapping of \textit{conv} layers as \acp{gemm} can also be performed using a relaxed form of the Toeplitz matrix~\cite{Bareiss1969}. However, the downside for using \acp{gemm} for the  layers is the introduction of redundant data in the input \acp{fm}. This redundancy, as pointed-out in~\cite{Sze2017}, can lead to either inefficiency in storage or complex memory access patterns. As a result, other strategies to map convolutions are considered.
\begin{equation}
\label{eq:convGEMM}
\matr{\tilde{Y}}^{\text{conv}} = \matr{\tilde{\Theta}}^{\text{conv}} \times \matr{\tilde{X}}^{\text{conv}}
\end{equation}

\subsection{Winograd Transform}
Winograd minimal filter algorithm, introduced in~\cite{Winograd1980}, is a computational transform that can be applied to convolutions when the stride is 1. Winograd convolutions are particularly efficient when processing small convolutions ($K \leq 3$), as demonstrated in ~\cite{Lavin2015a}.  In this works, authors report an acceleration up to x7.28 when compared to classical \ac{gemm} based implementation of convolutions when executing VGG16 on a TitanX \ac{gpu}.

In Winograd filtering,  data is processed by blocs referred as \emph{tiles}, as following:
\begin{enumerate}
	\item An input \ac{fm} tile $x$ of size $(u \times u)$ is pre-processed:  $\tilde{x} = \matr{A}^T x \matr{A}$ 
	\item In a same way, $\theta$ the filter tile of size $(k \times k)$ is transformed into $\tilde{\theta}$: $\tilde{\theta} = \matr{B}^T x \matr{B}$
	\item Winograd filtering algorithm, denoted $F(u \times u , k \times k)$, outputs a tile $y$ of size $(u \times u)$ that is computed according to equation~\ref{WinogradEq:yTile}
\end{enumerate}
\begin{equation}
y = \matr{C}^T \left[ \tilde{\theta} \odot \tilde{x} \right] \matr{C} \label{WinogradEq:yTile}
\end{equation}
	where $\matr{A},\matr{B},\matr{C}$ are transformation matrices defined in the Winograd algorithm~\cite{Winograd1980} and $\odot$ denotes the Hadamard product or \ac{ewmm}.

% \begin{align}
% 	& \tilde{x} = \matr{A}^T x \matr{A} \label{WinogradEq:xTile}\\
% 	& \tilde{\theta} = \matr{B}^T x \matr{B} \label{WinogradEq:thetaTile}\\
% 	& y = \matr{C}^T \left[ \tilde{\theta} \odot \tilde{x} \right] \matr{C} \label{WinogradEq:yTile}
% \end{align}.

While a standard filtering requires $u^2 \times k^2$ multiplications,  Winograd algorithm $F(u \times u , k \times k)$ requires $(u+k-1)^2$ multiplications~\cite{Winograd1980}. In the case of tiles of a size $u=2$ and kernels of size $k=3$, this corresponds to an arithmetic complexity reduction of x2.25~\cite{Lavin2015a}. In return, the number of additions is increased.

Beside this complexity reduction, implementing Winograd filtering in \ac{fpga}-Based \ac{cnn} accelerators has two advantages. First, transformation matrices $\matr{A},\matr{B},\matr{C}$ can be generated off-line once $u$ and $k$ are determined. As a result, these transforms become multiplications with the constants that can be implemented by means of \ac{lut} and shift registers, as proposed in~\cite{Lu2017}.

Second, Winograd filtering can employ the loop optimization techniques discussed in section~\ref{sec:LoopOpt} to vectorize the implementation. On one hand, the computational throughput is increased when \emph{unrolling} the computation of the \acp{ewmm} parts on an array of \ac{dsp} blocs. On the other hand, memory bandwidth is optimized using loop \emph{tiling} to determine the size \ac{fm} tiles and filter buffers.

First utilization of Winograd filtering in \ac{fpga}-Based \ac{cnn} accelerators is proposed in~\cite{DiCecco2016} and delivers a computational throughput of 46 GOPs when executing AlexNet convolution layers. This performance is significantly  by a factor of x42 in~\cite{Aydonat2017} when optimizing the datapath to support Winograd convolutions (by employing loop unrolling and tiling strategies), and storing the intermediate \ac{fm} in on-chip buffers~(cf sec~\ref{LoopOpti}). The same methodology is employed in~\cite{Lu2017} to derive a \ac{cnn} accelerator on a Xilinx ZCU102 device. This accelerator delivers a throughput of 2.94 TOPs on VGG convolutional layers, which corresponds to half of the performance of a TitanX device, with x5.7 less power consumption~\cite{Nvidia2015}\footnote{Implementation in the TitanX GPU employs Winograd algorithm and 32 bits floating point arithmetic}.

\subsection{Fast Fourier Transform}
%\Plan{Studied in~\cite{Lavin2015a}. Implemented on \acp{fpga}  in~\cite{Ko2017,Zhang2017}}.\\
\acf{fft} is a well known algorithm to transform the 2D convolutions into \ac{ewmm} in the frequency domain, as shown in equation~\ref{fft}:
\begin{equation}
	\label{fft}
	\text{\textbf{conv2D}}(X[c],\Theta[n,c]) = \text{IFFT} \Big(\text{FFT}(X[c]) \odot \text{FFT}(\Theta[n,c]) \Big)
\end{equation}

Using \ac{fft} to process 2D convolutions reduces the arithmetic complexity to $O(W^2 log_2(W))$, which is exploited to derive \ac{fpga}-based accelerators to \emph{train} \acp{cnn}~\cite{Ko2017}. When compared to standard filtering and Winograd algorithm, \ac{fft} finds its interest in convolutions with large kernel size $(K > 5)$, as demonstrated in~\cite{Lavin2015a,Bottleson2016}. The computational complexity of FFT convolutions can be further reduced to $O(W log_2(K))$ using the Overlap-and-Add Method~\cite{Smith1997} that can be applied when the signal size is much larger than the filter size, which is the case in \textit{conv} layers ($W >> K$). Works in~\cite{Zhang2017} exploit this method to implement frequency domain acceleration for \textit{conv} layers on \ac{fpga}, which results in a computational throughput of 83 GOPs for AlexNet.
%Works in~\cite{Bottleson2016,Nurvitadhi2017} investigate the two stategies.
\begin{table}
\caption{\ac{fpga}-Based \ac{cnn} accelerators employing computational transform to accelerate conv layers}
\label{TransformSurvey}
\resizebox{\textwidth}{!}{
% Table generated by Excel2LaTeX from sheet 'Transforms'
\begin{tabular}{|c|c|c|c|c|c|c|c|c|c|c|c|c|c|}
\cmidrule{3-14}\multicolumn{1}{c}{\multirow{2}[4]{*}{}} & \multirow{2}[4]{*}{} & \multirow{2}[4]{*}{\textbf{Network}} & \multicolumn{2}{c|}{\textbf{Network Workload}} & \multirow{2}[4]{*}{\textbf{Bitwidth}} & \multirow{2}[4]{*}{\textbf{Desc.}} & \multirow{2}[4]{*}{\textbf{Device}} & \textbf{Freq} & \textbf{Through} & \textbf{Power} & \textbf{LUT} & \multirow{2}[4]{*}{\textbf{DSP}} & \textbf{Memory} \\
\cmidrule{4-5}\multicolumn{1}{c}{} &       &       & \textbf{Comp. (GOP)} & \textbf{Param. (M)} &       &       &       & \textbf{(MHz)} & \textbf{(GOPs)} & \textbf{(W)} & \textbf{(K)} &       & \textbf{(MB)} \\
\midrule
\multirow{4}[8]{*}{Winograd} & \cite{DiCecco2016} & AlexNet-C & 1.3   & 2.3   & Float 32 & OpenCL & Virtex7 VX690T & 200   & 46    &       & 505   & 3683  & 56.3 \\
\cmidrule{2-14}      & \cite{Aydonat2017} & AlexNet-C & 1.3   & 2.3   & Float16 & OpenCL & Arria10 GX1150 & 303   & 1382  & 44.3  & 246   & 1576  & 49.7 \\
\cmidrule{2-14}      & \multirow{2}[4]{*}{\cite{Lu2017}} & VGG16-C & 30.7  & 14.7  & \multirow{2}[4]{*}{Fixed 16} & \multirow{2}[4]{*}{HLS} & \multirow{2}[4]{*}{Zynq ZU9EG} & \multirow{2}[4]{*}{200} & 3045  & \multirow{2}[4]{*}{23.6} & \multirow{2}[4]{*}{600} & \multirow{2}[4]{*}{2520} & \multirow{2}[4]{*}{32.8} \\
\cmidrule{3-5}\cmidrule{10-10}      &       & AlexNet-C & 1.3   & 2.3   &       &       &       &       & 855   &       &       &       &  \\
\midrule
\multirow{2}[4]{*}{FFT} & \multirow{2}[4]{*}{\cite{Zhang2017}} & AlexNet-C & 1.3   & 2.3   & \multirow{2}[4]{*}{Float 32} & \multirow{2}[4]{*}{} & \multirow{2}[4]{*}{Stratix5 QPI } & \multirow{2}[4]{*}{200} & 83    & \multirow{2}[4]{*}{13.2} & \multirow{2}[4]{*}{201} & \multirow{2}[4]{*}{224} & \multirow{2}[4]{*}{4.0} \\
\cmidrule{3-5}\cmidrule{10-10}      &       & VGG19-C & 30.6  & 14.7  &       &       &       &       & 123   &       &       &       &  \\
\midrule
\multirow{5}[10]{*}{GEMM} & \cite{Suda2016} & AlexNet-C & 1.3   & 2.3   & Fixed 16 & OpenCL & Stratix5 GXA7 & 194   & 66    & 33.9  & 228   & 256   & 37.9 \\
\cmidrule{2-14}      & \multirow{2}[4]{*}{\cite{Zhang2016a}} & \multirow{2}[4]{*}{VGG16-F} & \multirow{2}[4]{*}{31.1} & \multirow{2}[4]{*}{138.0} & \multirow{2}[4]{*}{Fixed 16} & \multirow{2}[4]{*}{HLS} & Kintex KU060 & 200   & 365   & 25.0  & 150   & 1058  & 14.1 \\
\cmidrule{8-14}      &       &       &       &       &       &       & Virtex7 VX960T & 150   & 354   & 26.0  & 351   & 2833  & 22.5 \\
\cmidrule{2-14}      & \multirow{2}[4]{*}{\cite{Zhang2017a}} & \multirow{2}[4]{*}{VGG16-F} & \multirow{2}[4]{*}{31.1} & \multirow{2}[4]{*}{138.0} & Fixed 16 & OpenCL & \multirow{2}[4]{*}{Arria10 GX1150} & 370   & 866   & 41.7  & 437   & 1320  & 25.0 \\
\cmidrule{6-7}\cmidrule{9-14}      &       &       &       &       & Float 32 & OpenCL &       & 385   & 1790  & 37.5  &       & 2756  & 29.0 \\
\bottomrule
\end{tabular}%

}
\end{table}

    \section{Data-path Optimizations for FPGA-Based CNN Accelerators}
\label{sec:OptDP}
% \begin{itemize}
% \item \Plan{Processing of \acp{cnn} exhibit large parallelism}
% \item \Plan{Given the available resource on \acp{fpga}, it impossible to fully pipeline and vectorize the hole processing, even for a layer}
% \item \Plan{Solution of SotA is to map a limited number of processing elements PEs are reused by iterating the data through them in software}
% \item \Plan{Early implementations relyed on Systolic arrays~\cite{Sankaradas2009,Farabet2009b,Chakradhar2010,Farabet2012,Gokhale2014} as they were efficient at 2D convolution filetering. However, systolic arrays suffers from complex routing logic and are inflexible as they can support convolutions only up to the maximum implemented kernel~\cite{Peemen2013}}
% \item \Plan{The problem boils dows to finding the best combination of (Architectural PE configuration + temporal scheduling) that maximizes throughput, or minimize energy consumption.}
% \item \Plan{This problem is now addressed by Loop-tiling and unrolling~\cite{Zhang2015}}
% \item \Plan{datapath is configured in adequation with the computational transform, \cite{Suda2016,Zhang2016} design OpenCL based kernels to accelerate GEMMs while \cite{Aydonat2017,Lu2017} optimize PE to Winograd based convolutions}
% \end{itemize}

As highlighted in sec~\ref{Workload:parallelism}, the execution of \acp{cnn} exhibit numerous sources of parallelism. However, because of the resource limitation of \acp{fpga} devices,  it is impossible to fully exploit all the parallelism patterns, especially with the sheer volume of operations involved in deep topologies. In other words, the execution of recent \ac{cnn} models can not fully be "Unrolled", sometimes, not even for a single \emph{conv} layer. To address this problem, the main approach that state-of-the-art implementations advocates, is to map a limited number of \acp{pe} on the \ac{fpga}. These \acp{pe} are reused by temporally iterating data through them. 
\begin{figure}[ht]
    \centering
    \subfloat[Static Systolic Array]{\label{img:SystolicArray}\includegraphics[width=0.3\textwidth]{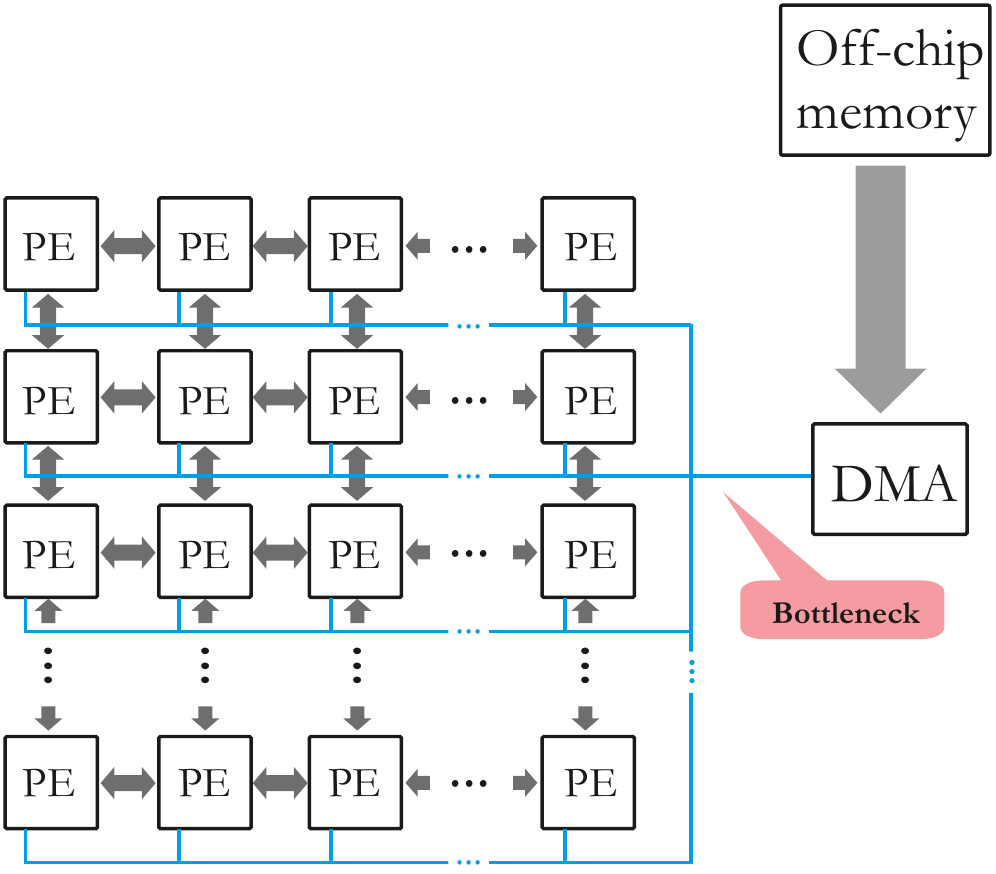}}
    \subfloat[Generic SIMD Accelerator]{\label{GenAccelerator}	 \includegraphics[width=0.3\textwidth]{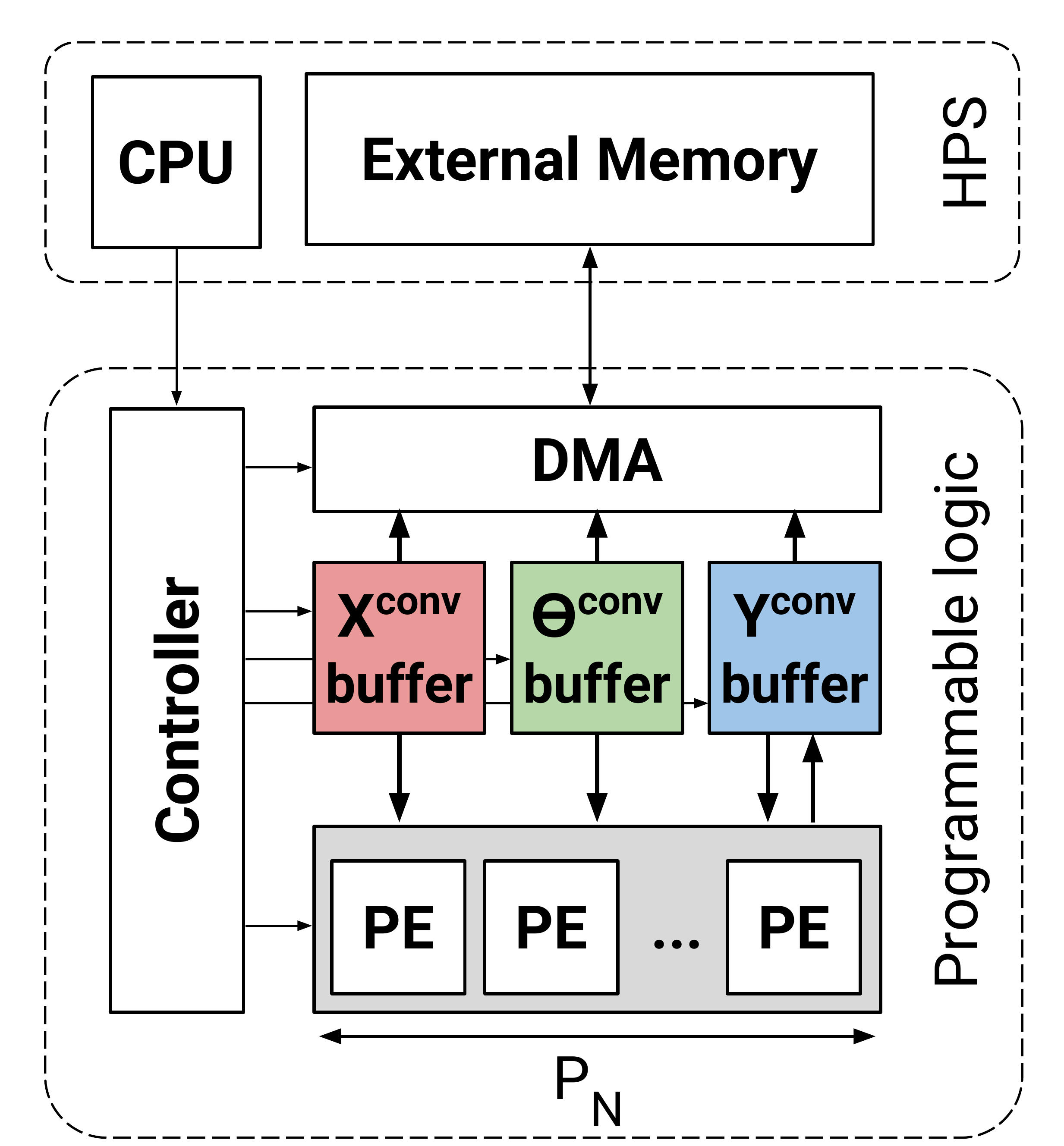}}
    \subfloat[Processing Element]{\label{ArchiPE}			 \includegraphics[width=0.3\textwidth]{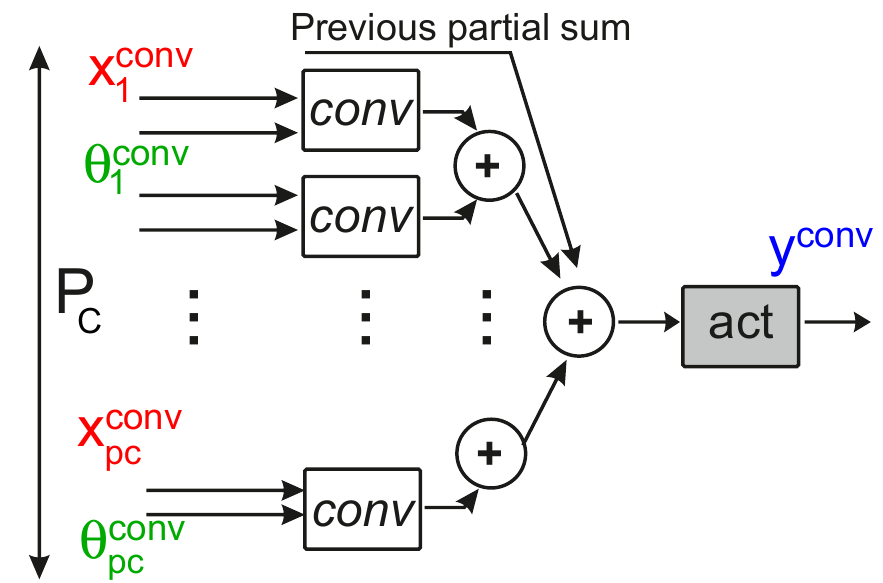}}
\caption{Generic Data-paths of FPGA-based CNN accelerators }
\end{figure}

\subsection{Systolic Arrays}
\label{sec:SystolicArrays}
Early \ac{fpga}-based accelerators for \acp{cnn} implemented systolic arrays to accelerate the 2D filtering in convolutions layers~\cite{Sankaradas2009,Farabet2009b,Chakradhar2010,Farabet2012,Gokhale2014}. As illustrated in figure~\ref{img:SystolicArray}, systolic arrays employ \emph{a static collection} of \acp{pe}, typically arranged in a 2-dimensional grid, that operates under the control of a CPU.
This static collection of PEs is agnostic to the CNN model configuration. It can only support convolutions with a kernel size $K$ that is smaller than a given maximum size $K_m$ (i.e support only convolutions such $K \leq K_m$ where , for instance, $K_m =7$ in~\cite{Farabet2009b} and $K_m = 10$ in~\cite{Gokhale2014}). Moreover, when performing convolutions with a smaller kernel size then $K_m$ ($K<<K_m$), only a small part of computing capabilities is used. For instance in~\cite{Gokhale2014}, processing $3 \times 3$ convolutions uses only 9\% of DSP Blocs. 
Finally, these systolic arrays do not implement data caching and requires to fetch inputs from off-chip memory. As a result, their performance is bounded by memory bandwidth of the device.

\subsection{SIMD Accelerators and Loop Optimization}
\label{sec:LoopOpt}
Due to inefficiency of static systolic arrays, flexible \ac{simd} accelerators for CNNs on FPGAs were proposed.   The general computation flow in these accelerators --illustrated in Fig.\ref{ArchiPE}-a-- is to fetch \acp{fm} and weights from \ac{dram} to on-chip buffers. These data are then streamed into the \acp{pe}. At the end of the \ac{pe} computation, results are transferred back to on-chip buffers and, if necessary, to the external memory in order to be fetched in their turn to process the next layers.
Each \ac{pe} --as depicted in Fig.~\ref{ArchiPE}-b-- is configurable and has its own \emph{computational} capabilities by means of \ac{dsp} blocs, and its own data \emph{caching} capabilities by means of on-chip registers.

With this paradigm, the problem of \ac{cnn} mapping boils down to finding the optimal architectural configuration of \acp{pe} (number of PEs, number of DSP blocs per PE, size of data caches), as well as the optimal temporal scheduling of data that maximizes the computational throughput $\mathscr{T}$. 
% \begin{figure}
% 	\includegraphics[width=.8\textwidth]{img/ArchiPE.png}
% 	\caption{\hl{Image from~\cite{Qiu2016} Re-draw me please, add numbers to execution cycles}}
% 	\label{ArchiPE}
% \end{figure}

For convolution layers, in which the processing is described in listing~\ref{convLayerLoop}, finding the optimal PE configuration can be seen as a loop optimization problem~\cite{Zhang2015,Qiu2016,Suda2016}
\cite{Atul2016,Zhang2016,Motamedi2016,Ma2016,Li2016a,Alwani2016,Ma2017,Wei2017}. This problem is addressed by applying loop optimization techniques such \emph{loop unrolling}, \emph{loop tiling} or \emph{loop interchange} to the 7 nested loops of listing~\ref{convLayerLoop}. In this case, setting the unroll and tiling factors (\textit{resp.} $P_i$ and $T_i$) determines the number of \acp{pe}, the computational resources and on-chip memory allocated to each \ac{pe} in addition to the size of on-chip buffer and the amount of \ac{dram} accesses.
\begin{table}
\centering
\caption{Loop Optimization Parameters $P_i$ and $T_i$}
\label{LoopOpti}
\begin{tabular}{|c|c|c|c|c|c|c|c|}
\hline
Parallelism  & Intra-layer & Inter-FM & \multicolumn{2}{c|}{Intra-FM} & Inter-Convolution & \multicolumn{2}{c|}{Intra-Convolution} \\ \hline
Loop & $L_L$       & $L_N$    & $L_V$         & $L_U$         & $L_C$             & $L_J$              & $L_K$             \\ \hline
Unroll factor         & $P_L$       & $P_N$    & $P_V$         & $P_U$         & $P_C$             & $P_J$              & $P_K$             \\ \hline
Tiling Factor         & $T_L$       & $T_N$    & $T_U$         & $T_U$         & $T_C$             & $T_J$              & $T_K$             \\ \hline
\end{tabular}
\end{table}

\subsubsection{Loop Unrolling:\\}
Unrolling a loop $L_i$ with an unrolling factor $P_i$ ($P_i \leq i, i \in \{L,V,U,N,C,J,K\}$) accelerates its execution at the expense of resource utilization. Each of the parallelism patterns listed in section~\ref{Workload:parallelism} can be implemented by unrolling one of the loops of listing~\ref{convLayerLoop}, as summarized in table~\ref{LoopOpti}. For configuration given in figure~\ref{ArchiPE}, the unrolling factor $P_N$ determines the number of \acp{pe}. On the other hand, unrolling factors $P_C, P_K , P_J$ determine the number of multipliers and adders, as well as the size of registers contained in each \ac{pe}.

\begin{figure}
	\centering
	\subfloat[]{\includegraphics[width=.38\textwidth]{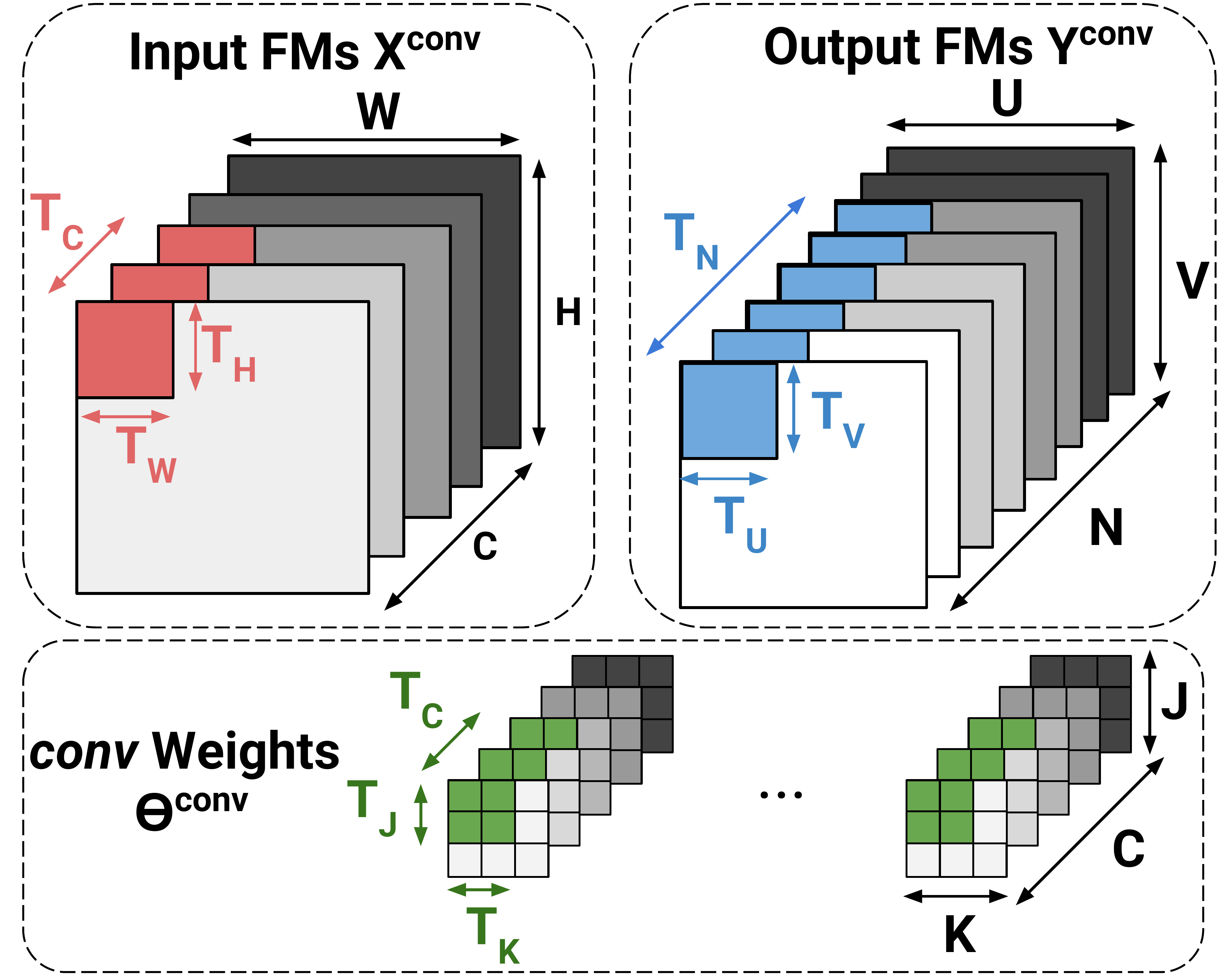}}
    \hfill
	\subfloat[]{\includegraphics[width=.5\textwidth]{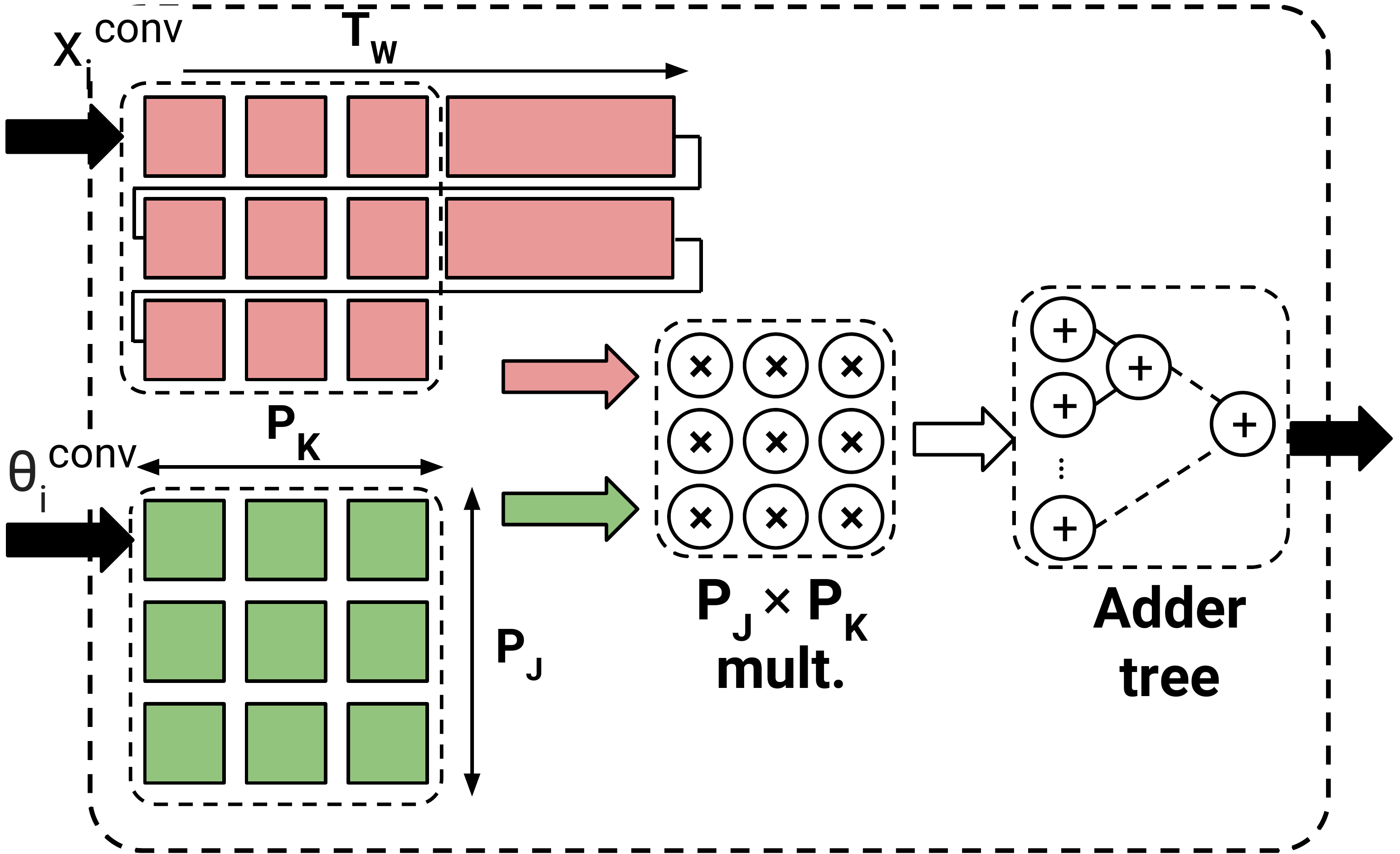}}
	\caption{{Loop tiling and unrolling}}
	\label{img:LoopOpt}
\end{figure}

\subsubsection{Loop Tiling:\\}
% \begin{itemize}
% 	\item \Plan{Define loop tiling}
%     \item \Plan{$1 < P_i < T_i < i$}
%     \item \Plan{Value of $T$ controls the capacity of on-chip buffer}
%     \item \Plan{\cite{Zhang2015} Tiles loops $L_C,L_N$. In addition $ T_C= P_C$ and $ T_N = P_N$, more performance when using \ac{fpga} cluster as shown in \cite{Zhang2016}}
%     \item \Plan{\cite{Motamedi2016}   expands tiling to loops $L_K$ . $ T_K= P_K$}
%     \item \Plan{\cite{Atul2016} tiles across $T_U,T_V, T_N$ }
%     \item \Plan{To keep data in on-chip buffer after execution of layer. \cite{Alwani2016} investigates Fused-Layer \ac{cnn} Accelerators$T_L < L$}
%     \item \Plan{Our saviour, lord~\cite{Ma2017} separates unrolling and tiling and fully explores the design space}
% 	\item \Plan{His result in\cite{Ma2017} : Only the computation within one input feature map and across multiple output feature maps are unrolled. By this means, both pixels and weights are reused by multiple PEs and high level of parallelism can be supported with large}
% 	\item \Plan {$P_K = P_J = P_C = 1$}
% 	\item \Plan {$P_U > 1, P_V > 1, P_N > 1$}
% \end{itemize}
In general, the capacity of on-chip memory in current \acp{fpga} is not large enough to store all the weights and intermediate \acp{fm} of all \ac{cnn} layers%\footnote{Exception can be made for \cite{Burger17}, where a large cluster of \acp{fpga} is deployed and resorts only to on-chip memory to store \ac{cnn} weights and data}.
As a consequence, \ac{fpga} based accelerators resort to external \acp{dram} to store this data. As mentioned in section~\ref{Workload:Memory}, \ac{dram} accesses are costly in terms of energy and latency, and data caches must be implemented by means of on-chip buffers and local registers. The challenge is to configure the data-path in a way that every data transferred from \ac{dram} is reused as much as possible.

For \textit{conv} layers, this challenge can be addressed by \emph{tiling} the nested loops of listing~\ref{convLayerLoop}. \emph{Loop tiling}~\cite{Derrien2001} divides the \acp{fm} and weights of each layer into multiple blocks that can fit into the on-chip buffers. For the configuration given in figure~\ref{ArchiPE}, sizes of buffers containing input \ac{fm}, weights and output \ac{fm} are determined by the tiling factors detailed in table~\ref{LoopOpti}, according to equation~\ref{eq:BufferSize}%s \ref{BufferX},\ref{BufferTheta} and \ref{BufferY}.
% \begin{align}
% &\mathscr{M}_{X}^{\text{conv}} = T_C \times T_H \times T_W \label{BufferX}\\
% &\mathscr{M}_{\Theta}^{\text{conv}} = T_N \times T_C \times T_J \times T_K \label{BufferTheta}\\
% &\mathscr{M}_{Y}^{\text{conv}} = T_N \times T_V \times T_U \label{BufferY}
% \end{align}
\begin{equation}
\label{eq:BufferSize}
\mathscr{M}_{\text{conv}} = T_C T_H T_W + T_N T_C T_J T_K + T_N T_V T_U
\end{equation}
% For each loop $L_i$, tiling factors have a value between $i$ and $P_i$, as shown in equation~\ref{LoopOptiVal}
% \begin{align}
% \nonumber
% \forall i & \in \{L,N,C,U,V,J,K \} \\
% & 1 \leq P_i \leq T_i \leq i \label{LoopOptiVal}
% \end{align}

% Thus, \ac{fpga} accelerators for \ac{cnn} employ loop tiling to address the bandwidth bottleneck problem. This is the case in~\cite{Zhang2015,Zhang2016,Ma2016}, where authors tile loops $L_N,L_C$ to accelerate the execution of \textit{conv} layers.
% Implementations in \cite{Suda2016,Qiu2016,Li2016a,Motamedi2016} expand the tiling to loops $L_J,L_K$ in order to reuse the kernels for each convolution. This tiling requires \acp{pe} to be configured differently for different layers, increasing thus the control complexity.  Finaly, in order to keep data in on-chip buffer after execution of a given layer,\cite{Alwani2016} investigates fused-layer \ac{cnn} Accelerators by tiling across layer $L_L$.

% \begin{minipage}{\linewidth}
% 	\lstinputlisting[language=mycpp,label=convLayerTiled,caption={Tiling Loops $L_N,L_C,L_V,L_U$} in conv layers]{listing/convLayerTiled.cpp}
% \end{minipage}

\begin{figure}[ht]
\subfloat[]{
	\begin{minipage}{0.4\textwidth}
\begin{lstlisting}[language=mycpp,label=convLayerLoop]
// Ll: Layer
for (int l=0;l<L,l++){
// Lb : Batch
for (int b=0;b<B,l++){
// Ln: Y Depth
for (int n=0;n<N;n++){
// Lv: Y Columns
for (int v=0;v<V,v++){
// Lu: Y Raws
for (int u=0;u<U,u++){
// Lc: X Depth
for (int c=0;n<C;c++){
// Lj: Theta Columns
for (int j=0;j<J,j++){
// Lk: Theta Raws
for (int k=0;k<K,k++){ 
  Y[b,l,n,v,u] += X[b,l,c,v+j,u+k] * Theta[l,n,c,j,k]
}}}}}}}
\end{lstlisting}
	\end{minipage}
}
\hfill
\subfloat[]{
	\begin{minipage}{0.55\textwidth}
\begin{lstlisting}[language=mycpp,label=convLayerTiled]
for (int n=0;n<N;n+=Tn){ 
for (int v=0;v<V,v+=Tv){ 
for (int u=0;u<U,u+=Tu){ 
for (int c=0;n<C;c+=Tc){
  // DRAM: Load in on-chip buffers the tiles: 
  // X[l,c:c+Tc,v:v+Tv,u:u+Tu]
  // Theta [l,n:n+Tn,c:c+Tc,j,k]
  // Process on-chip tiles
  for (int tn=0;tn<Tn;tn++){ 
  for (int tv=0;tv<Tv,tv++){ 
  for (int tu=0;tu<Tu,tu++){ 
  for (int tc=0;tn<Tc;tc++){
  for (int j=0;j<J,j++){ 
  for (int k=0;k<K,k++){
    Y[l,tn,tv,tu] += X[l,tc,tv+j,tu+k] * Theta[l,tn,tc,j,k];
  }}}}}}
  // DRAM: Store output tile 
}}}}
\end{lstlisting}
    \end{minipage}
}
\caption{Loop Tiling in conv layers: a-Before tiling, b-After tiling}
\end{figure}

\subsubsection{Design Space Exploration:\\}
% \Plan{- Search for optimal $P_i,T_i$ that minimize the memory access and data movements while maximizing the resource utilization.}

In order to find the optimal unrolling and tiling factors, a large exploration of the design space is needed. In a general way, an analytical model is built. Inputs of this model are loop factors $P_i,T_i$ and outputs are a theoretical prediction of the allocated resources, the computational throughout  and the memory bandwidth used. This model is parametrized by the available resources of a given \ac{fpga} platform and the workload of the CNN.

Given this model, the objective is to find the design parameters that minimize the memory access while maximizing the resource utilization. To address this optimization problem, a brute force exploration is performed, such in~\cite{Zhang2015,Suda2016,Atul2016,Zhang2016,Motamedi2016,Ma2016}. This exploration is usually driven by the Roofline method~\cite{Williams2009} in order to select the feasible design solutions that matches with the maximum computational throughput and the maximum memory bandwidth a given platform can deliver~\cite{Zhang2015,Motamedi2016,Meloni2016}. The design space can also be explored by means of heuristic search algorithms, as proposed for instance in~\cite{Sharma2016}.

\subsubsection{\ac{fpga} Implementations:\\}
Employing loop optimizations to derive \ac{fpga}-based \ac{cnn} accelerator was first investigated in~\cite{Zhang2015}. In this work, Zhang \textit{et al.} report a computational throughput of 61.62 GOPs in the execution of AlexNet convolutional layers by unrolling loops $L_C$ and $L_N$.  This accelerator was built using HLS tools and rely on 32 floating point arithmetic. Works in~\cite{Ma2016} follow the same unrolling scheme and implement the \ac{fc} part of the inference. Moreover, design~\cite{Ma2016} features 16 bits fixed point arithmetic and RTL conception, resulting in a x2.2 improvement in terms of computational throughput. Finally, the same unrolling and tiling scheme are employed in recent works~\cite{Zhang2016} were authors report a x13,4 improvement over their original works in~\cite{Zhang2015}, thanks to a deeply pipelined \ac{fpga} cluster of four Virtex7-XV960t devices and a 16 bits fixed point arithmetic.

In all these implementations, loops $L_J$ and $L_K$ are not unrolled because $J$ and $K$ are usually small, especially in recent topologies (cf Table~\ref{tab:PopularCNNs}). Works of Motamedi \textit{et al.}~\cite{Motamedi2016} study the impact of unrolling these loops in AlexNet, where the first convolution layers use $11 \times 11$ and $ 5 \times 5$  filters. Expanding loop unrolling and tiling to loops $L_J$ and $L_K$ results in a x1.36 improvement in computational throughput \textit{vs} \cite{Zhang2015} on the same VX485T device when using 32 floating point arithmetic.
In a same way, implementations in~\cite{Suda2016,Qiu2016,Li2016a} tile and unroll loops $L_N,L_C,L_J,L_K$ and demonstrate higher acceleration on AlexNet and VGG when using fixed point arithmetic. Nevertheless, and as pointed out in~\cite{Ma2017}, unrolling loops  $L_J$ and $L_K$ is ineffective for recent \ac{cnn} models that employ small convolution kernels. In addition, Tiling loops $L_J$ and $L_K$ requires \acp{pe} to be configured differently for different layers, increasing thus the control complexity.

The values of $U,V,N$ can be very large in \ac{cnn} models. Consequently, unrolling and tiling loops $L_U,L_V,L_N$ can be efficient only for devices with high computational capabilities (i.e DSP Blocs). This is demonstrated in works of Rahman \textit{et al.}~\cite{Atul2016} that report an improvement of  $\times 1.22$ over~\cite{Zhang2015} when enlarging the design space exploration to loops $L_U,L_V,L_N$

In order to keep data in on-chip buffer after the execution of a given layer,\cite{Alwani2016} investigates fused-layer \ac{cnn} Accelerators by tiling across layer $L_L$. As a result, authors report a reduction of 95\% of DRAM accesses at the cost of 362KB of extra on-chip memory.

In all these approaches, loops $L_N,L_C,L_J,L_K$ are unrolled in a same way they are tilled (i.e $T_i = P_i$). By contrast, the works of Ma \textit{et al.}~\cite{Ma2017,Ma2017b} fully explore all the design variables searching for optimal loop unroll and tiling factors. More particularly, authors demonstrate that the input \acp{fm} and weights are optimally reused when unrolling only computations within a single input \ac{fm} (i.e when $P_C = P_J = P_k = 1$). Tiling factors are set in way that all the data required to compute an element of $Y$ are fully buffered (i.e $T_C = C, T_K = K, T_J = J$). The remaining design parameters are derived after a brute force design exploration. The same authors leverage on these loop optimizations to build an RTL compiler for \acp{cnn} in~\cite{Ma2017a}. To the best of our knowledge, this accelerator outperforms all the previous implementations that are based on loop optimization in terms of computational throughput.

\begin{table}
\caption{\ac{fpga}-based \ac{cnn} accelerators implementing loop optimization}
\label{LoopOptSurvey}
\resizebox{\textwidth}{!}{
% Table generated by Excel2LaTeX from sheet 'LoopOptimization'
\begin{tabular}{|c|c|c|c|c|c|c|c|c|c|c|c|c|}
\cmidrule{2-13}\multicolumn{1}{c|}{\multirow{2}[4]{*}{}} & \multirow{2}[4]{*}{\textbf{Network}} & \multicolumn{2}{c|}{\textbf{Network Workload}} & \multirow{2}[4]{*}{\textbf{Bitwidth}} & \multirow{2}[4]{*}{\textbf{Desc.}} & \multirow{2}[4]{*}{\textbf{Device}} & \textbf{Freq} & \textbf{Through} & \textbf{Power} & \textbf{LUT} & \multirow{2}[4]{*}{\textbf{DSP}} & \textbf{Memory} \\
\cmidrule{3-4}\multicolumn{1}{c|}{} &       & \textbf{Comp. (GOP)} & \textbf{Param. (M)} &       &       &       & \textbf{(MHz)} & \textbf{(GOPs)} & \textbf{(W)} & \textbf{(K)} &       & \textbf{(MB)} \\
\midrule
\cite{Zhang2015} & AlexNet-C & 1.3   & 2.3   & Float 32 & HLS   & Virtex7 VX485T & 100   & 61.62 & 18.61 & 186   & 2240  & 18.4 \\
\midrule
\cite{Qiu2016} & VGG16SVD-F & 30.8  & 50.2  & Fixed 16 & HDL   & Zynq Z7045 & 150   & 136.97 & 9.63  & 183   & 780   & 17.5 \\
\midrule
\multirow{3}[6]{*}{\cite{Suda2016}} & AlexNet-C & 1.3   & 2.3   & \multirow{3}[6]{*}{Fixed 16} & \multirow{3}[6]{*}{OpenCL} & \multirow{3}[6]{*}{Stratix5 GSD8} & \multirow{3}[6]{*}{120} & 187.24 & \multirow{4}[8]{*}{33.93} & 138   & 635   & 18.2 \\
\cmidrule{2-4}\cmidrule{9-9}\cmidrule{11-13}      & AlexNet-F & 1.4   & 61.0  &       &       &       &       & 71.64 &       & 272   & 752   & 30.1 \\
\cmidrule{2-4}\cmidrule{9-9}\cmidrule{11-13}      & VGG16-F & 31.1  & 138.0 &       &       &       &       & 117.9 &       & 524   & 1963  & 51.4 \\
\cmidrule{1-9}\cmidrule{11-13}\cite{Atul2016} & AlexNet-C & 1.3   & 2.3   & Float 32 & HLS   & Virtex7 VX485T & 100   & 75.16 &       & 28    & 2695  & 19.5 \\
\midrule
\multirow{2}[4]{*}{\cite{Zhang2016}} & AlexNet-F & 1.4   & 61.0  & \multirow{2}[4]{*}{Fixed 16} & \multirow{2}[4]{*}{HLS} & \multirow{2}[4]{*}{Virtex7 VX690T} & \multirow{2}[4]{*}{150} & 825.6 & 126.00 &       & 14400 &  \\
\cmidrule{2-4}\cmidrule{9-13}      & VGG16-F & 31.1  & 138.0 &       &       &       &       & 1280.3 & 160.00 &       & 21600 &  \\
\midrule
\multirow{2}[4]{*}{\cite{Ma2016}} & NIN-F & 2.2   & 61.0  & \multirow{2}[4]{*}{Fixed 16} & \multirow{2}[4]{*}{HDL} & \multirow{2}[4]{*}{Stratix5 GXA7} & \multirow{2}[4]{*}{100} & 114.5 & 19.50 & 224   & 256   & 46.6 \\
\cmidrule{2-4}\cmidrule{9-13}      & AlexNet-F & 1.5   & 7.6   &       &       &       &       & 134.1 & 19.10 & 242   & 256   & 31.0 \\
\midrule
\cite{Li2016a} & AlexNet-F & 1.4   & 61.0  & Fixed 16 &       & Virtex7 VX690T & 156   & 565.94 & 30.20 & 274   & 2144  & 34.8 \\
\midrule
\cite{Alwani2016} & AlexNet-C & 1.3   & 2.3   & Float 32 & HLS   & Virtex7 VX690T & 100   & 61.62 &       & 273   & 2401  & 20.2 \\
\midrule
\cite{Ma2017} & VGG16-F & 31.1  & 138.0 & Fixed 16 & HDL   & Arria10 GX1150 & 150   & 645.25 & 50.00 & 322   & 1518  & 38.0 \\
\midrule
\multirow{3}[6]{*}{\cite{Wei2017}} & AlexNet-F & 1.4   & 61.0  & \multirow{3}[6]{*}{Fixed 16} & \multirow{3}[6]{*}{OpenCL} & \multirow{3}[6]{*}{Arria10 GT1150} & 239.6 & 360.4 & \multirow{5}[10]{*}{} & 700   & 1290  & 47.2 \\
\cmidrule{2-4}\cmidrule{8-9}\cmidrule{11-13}      & VGG-F & 31.1  & 138.0 &       &       &       & 221.65 & 460.5 &       & 708   & 1340  & 49.3 \\
\cmidrule{2-4}\cmidrule{8-9}\cmidrule{11-13}      & VGG-F & 31.1  & 138.0 &       &       &       & 231.85 & 1171.3 &       & 626   & 1500  & 33.4 \\
\cmidrule{1-9}\cmidrule{11-13}\multirow{2}[4]{*}{\cite{Motamedi2017}} & \multirow{2}[4]{*}{AlexNet-C} & 1.3   & 2.3   & \multirow{2}[4]{*}{Fixed 16} & \multirow{2}[4]{*}{HDL} & Cyclone5  SEM & 100   & 12.11 &       & 22    & 28    & 0.2 \\
\cmidrule{3-4}\cmidrule{7-9}\cmidrule{11-13}      &       & 1.3   & 2.3   &       &       & Virtex7 VX485T & 100   & 445   &       &       & 2800  &  \\
\midrule
\multirow{6}[12]{*}{\cite{Ma2017a}} & NiN   & 20.2  & 7.6   & \multirow{6}[12]{*}{Fixed 16} & \multirow{6}[12]{*}{HDL} & \multirow{3}[6]{*}{Stratix5 GXA7} & \multirow{3}[6]{*}{150} & 282.67 & \multirow{6}[12]{*}{} & 453   & 256   & 30.2 \\
\cmidrule{2-4}\cmidrule{9-9}\cmidrule{11-13}      & VGG16-F & 31.1  & 138.0 &       &       &       &       & 352.24 &       & 424   & 256   & 44.0 \\
\cmidrule{2-4}\cmidrule{9-9}\cmidrule{11-13}      & ResNet-50 & 7.8   & 25.5  &       &       &       &       & 250.75 &       & 347   & 256   & 39.3 \\
\cmidrule{2-4}\cmidrule{7-9}\cmidrule{11-13}      & NiN   & 20.2  & 7.6   &       &       & \multirow{3}[6]{*}{Arria10 GX1150} & \multirow{3}[6]{*}{200} & 587.63 &       & 320   & 1518  & 30.4 \\
\cmidrule{2-4}\cmidrule{9-9}\cmidrule{11-13}      & VGG16-F & 31.1  & 138.0 &       &       &       &       & 720.15 &       & 263   & 1518  & 44.5 \\
\cmidrule{2-4}\cmidrule{9-9}\cmidrule{11-13}      & ResNet-50 & 7.8   & 25.5  &       &       &       &       & 619.13 &       & 437   & 1518  & 38.5 \\
\midrule
\multirow{2}[4]{*}{\cite{Liu2017}} & AlexNet-F & 1.5   & 7.6   & \multirow{2}[4]{*}{Float 32} & \multirow{2}[4]{*}{} & \multirow{2}[4]{*}{Virtex7 VX690T} & \multirow{2}[4]{*}{100} & 445.6 & 24.80 & 207   & 2872  & 37 \\
\cmidrule{2-4}\cmidrule{9-13}      & VGG16SVD-F & 30.8  & 50.2  &       &       &       &       & 473.4 & 25.60 & 224   & 2950  & 47 \\
\bottomrule
\end{tabular}%

}
\end{table}

\subsection{Dataflow MoC For CNNs}
% \begin{itemize}
% 	\item \Plan{\ac{cnn} graphs can be modelled as DPN}
% 	\item \Plan{Intermediate representation between \ac{cnn} and Hardware}
% 	\item \Plan{Design space exploration in this intermediate representation, using graph transforms}
% 	\item \Plan{\cite{Sharma2016,Venieris2016,Venieris2017}}
% \end{itemize}
Feed-forward propagation is by nature a streaming based applications in which the execution is purely data-driven. In fact, the \ac{cnn} layout is in contrast with Von Neumann execution models and a \ac{cnn} implementation can easily be memory-bounded if it has to fetch every instruction from memory. This motivated multiple approaches to investigate the applicability of the data-flow \ac{moc} to accelerate \acp{cnn} on \acp{fpga}.

The foundations of the data-flow \acp{moc} were formalized by~\cite{Dennis1974} in order to create an architecture where multiple fragments of instructions can process simultaneously streams of data. Programs respecting dataflow semantics are described as \acp{dpn}. Each node of this network corresponds to a fundamental processing unit called an \textit{actor} and each edge corresponds to a communication \textit{FIFO} channel. Actors exchange abstract data --known as \textit{tokens}-- through these FIFOs. Each actor follows a purely data-driven execution model wherein the \textit{firing} (execution) is triggered only by the availability of input operands. This is typically the case in \acp{cnn}, where the execution of each layer is only triggered by the availability of input \ac{fm}.

Applying the data-flow \ac{moc} to accelerate \ac{cnn} implementations on \acp{fpga} is investigated in~\cite{Li2016b}. In this work, authors demonstrate the efficiency of the proposed \textit{lightweight data-flow methodology}~\cite{Shen2010} by mapping \textit{conv} layers with variable clock-domains in a Zynq ZC706 device.

A special case of data-flow, referred as \ac{sdf}~\cite{Lee1987}, is a paradigm in which the number of tokens produced and consumed by each actor can be specified a priori, as it is the case in the CNN execution. \ac{sdf} model is employed in~\cite{Venieris2016,Venieris2017} to optimize the mapping of \ac{cnn} graphs on \acp{fpga}. In this works, the \ac{cnn} graph is modeled as a topology matrix that contains the the number of incoming streams, the size of tokens and the consumption rates of each actor. Instead of exploring the design space of unrolling and tiling parameters (cf. sec~\ref{sec:LoopOpt}), authors explore the design space of the topology matrix components. These optimal components are used to derive the configuration of the \ac{pe} and buffers that either minimizes the computation latency or energy consumption. Moreover, and in contrast with classical implementations where data is streamed in and out of layers using off-chip data transfers, authors exploit partial dynamic reconfiguration of \acp{fpga} to process different layers.

Finally, works in~\cite{Abdelouahab2017} optimize the direct hardware mapping of \ac{cnn} graphs. In this approach, each actor of the \ac{dpn} is physically mapped on the device with its own specific instance, while each edge is mapped as a signal. As all the computations are unrolled, applicability of this method can rapidly be limited by the resource of the device or the size of the \ac{cnn}, preventing this approach from implementing deep models.
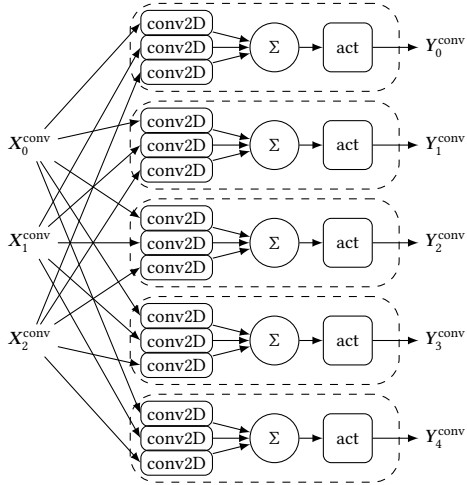
\begin{figure}[ht]
\centering
\begin{tikzpicture}[scale=0.65, every node/.style={scale=0.65}]

    \tikzset{neuron/.style={draw,circle,minimum size=1.2cm}};
    \tikzset{conv/.style={draw,rectangle,rounded corners=3pt,minimum size=0.4cm}};
    \tikzset{sum/.style={draw,circle,minimum size=1cm}};
    \tikzset{act/.style={draw,rectangle,rounded corners=3pt,minimum size=1cm}};
    \tikzset{neuronBox/.style={draw,rectangle,dashed,rounded corners=10pt,minimum width=5.5cm,minimum height=1.8cm}};
    \tikzset{void/.style={}}

    \node[void]     (i0)      at(-3,2)   {$\matr{X}^{ \text{conv}}_2$};
    \node[void]     (i1)      at(-3,4)   {$\matr{X}^{ \text{conv}}_1$};
    \node[void]     (i2)      at(-3,6)   {$\matr{X}^{ \text{conv}}_0$};
%    \node[]         (r)       at(2,4)    {\huge{\textcolor{red}{ADD NETLIST-VIEW}}};

    \node [conv] (ce0)  at (0,-0.5)   {{$\text{conv2D}$}};
    \node [conv] (ce1)  at (0,0)       {{$\text{conv2D}$}};
    \node [conv] (ce2)  at (0,0.5)    {{$\text{conv2D}$}};
    \node [conv] (ce3)  at (0,1.5)    {{$\text{conv2D}$}};
    \node [conv] (ce4)  at (0,2)       {{$\text{conv2D}$}};
    \node [conv] (ce5)  at (0,2.5)    {{$\text{conv2D}$}};
    \node [conv] (ce6)  at (0,3.5)    {{$\text{conv2D}$}};
    \node [conv] (ce7)  at (0,4)       {{$\text{conv2D}$}};
    \node [conv] (ce8)  at (0,4.5)    {{$\text{conv2D}$}};
    \node [conv] (ce9)  at (0,5.5)    {{$\text{conv2D}$}};
    \node [conv] (ce10) at (0,6)      {{$\text{conv2D}$}};
    \node [conv] (ce11) at (0,6.5)    {{$\text{conv2D}$}};
	\node [conv] (ce12) at (0,7.5)    {{$\text{conv2D}$}};
    \node [conv] (ce13) at (0,8)      {{$\text{conv2D}$}};
    \node [conv] (ce14) at (0,8.5)    {{$\text{conv2D}$}};
    
    \node[sum]   (sum0)      at(2,0)    {$\Sigma$};
    \node[sum]   (sum1)      at(2,2)    {$\Sigma$};
    \node[sum]   (sum2)      at(2,4)    {$\Sigma$};
    \node[sum]   (sum3)      at(2,6)    {$\Sigma$};
    \node[sum]   (sum4)      at(2,8)    {$\Sigma$};
    
    \node[act]   (act0)      at(3.5,0)    {$\mbox{act}$};
    \node[act]   (act1)      at(3.5,2)    {$\mbox{act}$};
    \node[act]   (act2)      at(3.5,4)    {$\mbox{act}$};
    \node[act]   (act3)      at(3.5,6)    {$\mbox{act}$};
    \node[act]   (act4)      at(3.5,8)    {$\mbox{act}$};

    \node[void]     (o0)      at(5.5,0)  {$\matr{Y}^{ \text{conv}}_4$};
    \node[void]     (o1)      at(5.5,2)  {$\matr{Y}^{ \text{conv}}_3$};
    \node[void]     (o2)      at(5.5,4)  {$\matr{Y}^{ \text{conv}}_2$};
    \node[void]     (o3)      at(5.5,6)  {$\matr{Y}^{ \text{conv}}_1$};
    \node[void]     (o4)      at(5.5,8)  {$\matr{Y}^{ \text{conv}}_0$};
    
    \node[neuronBox]   (n0)      at(1.8,0)  {};
    \node[neuronBox]   (n1)      at(1.8,2)  {};
    \node[neuronBox]   (n2)      at(1.8,4)  {};
    \node[neuronBox]   (n3)      at(1.8,6)  {};
    \node[neuronBox]   (n4)      at(1.8,8)  {};
    
    \draw[->,>=latex] (i0)--(ce0.west);
    \draw[->,>=latex] (i0)--(ce3.west);
    \draw[->,>=latex] (i0)--(ce6.west);
    \draw[->,>=latex] (i0)--(ce9.west);
    \draw[->,>=latex] (i0)--(ce12.west);

    \draw[->,>=latex] (i1)--(ce1.west);
    \draw[->,>=latex] (i1)--(ce4.west);
    \draw[->,>=latex] (i1)--(ce7.west);
    \draw[->,>=latex] (i1)--(ce10.west);
    \draw[->,>=latex] (i1)--(ce13.west);

    \draw[->,>=latex] (i2)--(ce2.west);
    \draw[->,>=latex] (i2)--(ce5.west);
    \draw[->,>=latex] (i2)--(ce8.west);
    \draw[->,>=latex] (i2)--(ce11.west);
    \draw[->,>=latex] (i2)--(ce14.west);

    \draw[->,>=latex] (ce0) --(sum0)   ;
    \draw[->,>=latex] (ce3) --(sum1)   ;
    \draw[->,>=latex] (ce6) --(sum2)   ;
    \draw[->,>=latex] (ce9) --(sum3)   ;
    \draw[->,>=latex] (ce12)--(sum4)   ;
    \draw[->,>=latex] (ce1) --(sum0)   ;
    \draw[->,>=latex] (ce4) --(sum1)   ;
    \draw[->,>=latex] (ce7) --(sum2)   ;
    \draw[->,>=latex] (ce10)--(sum3)   ;
    \draw[->,>=latex] (ce13)--(sum4)   ;
    \draw[->,>=latex] (ce2) --(sum0)   ;
    \draw[->,>=latex] (ce5) --(sum1)   ;
    \draw[->,>=latex] (ce8) --(sum2)   ;
    \draw[->,>=latex] (ce11)--(sum3)   ;
    \draw[->,>=latex] (ce14)--(sum4)   ;

    \draw[->,>=latex] (sum0)--(act0);
    \draw[->,>=latex] (sum1)--(act1);
    \draw[->,>=latex] (sum2)--(act2);
    \draw[->,>=latex] (sum3)--(act3);
    \draw[->,>=latex] (sum4)--(act4);

    \draw[->,>=latex] (act0)--(o0);
    \draw[->,>=latex] (act1)--(o1);
    \draw[->,>=latex] (act2)--(o2);
    \draw[->,>=latex] (act3)--(o3);
    \draw[->,>=latex] (act4)--(o4);
    
\end{tikzpicture}
\caption{An example of a graph representation of a convolution layer ($C=3,N=5$)}
\label{DPNConvLayer}
\end{figure}

\begin{table}
\caption{\ac{fpga}-Based \ac{cnn} accelerators employing the data-flow MoC}
\label{DataflowMOC}
\resizebox{\textwidth}{!}{
% Table generated by Excel2LaTeX from sheet 'DataflowMOC'
\begin{tabular}{|c|c|c|c|c|c|c|c|c|c|c|c|c|}
\cmidrule{2-13}\multicolumn{1}{c|}{\multirow{2}[4]{*}{}} & \multirow{2}[4]{*}{\textbf{Network}} & \multicolumn{2}{c|}{\textbf{Network Workload}} & \multirow{2}[4]{*}{\textbf{Bitwidth}} & \multirow{2}[4]{*}{\textbf{Desc.}} & \multirow{2}[4]{*}{\textbf{Device}} & \textbf{Freq} & \textbf{Through} & \textbf{Power} & \textbf{LUT} & \multirow{2}[4]{*}{\textbf{DSP}} & \textbf{Memory} \\
\cmidrule{3-4}\multicolumn{1}{c|}{} &       & \textbf{Comp. (GOP)} & \textbf{Param. (M)} &       &       &       & \textbf{(MHz)} & \textbf{(GOPs)} & \textbf{(W)} & \textbf{(K)} &       & \textbf{(KB)} \\
\midrule
\cite{Li2016} & CarType-C & 0.16  & 0.03  & Float 32 & HDL   & Zynq Z7045 & 100   & 0.47  & 0.23  & 68    & 24    & 1440.0 \\
\midrule
\multirow{2}[4]{*}{\cite{Venieris2016}} & LeNet5-C & 0.04  & 0.03  & \multirow{2}[4]{*}{Fixed 16} & \multirow{2}[4]{*}{HLS} & \multirow{2}[4]{*}{Zynq Z7020} & \multirow{2}[4]{*}{100} & 0.48  & \multirow{2}[4]{*}{0.75} & 14    & 4     & 42.7 \\
\cmidrule{2-4}\cmidrule{9-9}\cmidrule{11-13}      & SignRecog-C & 4.03  & 0.04  &       &       &       &       & 6.03  &       & 26    & 144   & 38.2 \\
\midrule
\cite{Venieris2017} & VGG16-F & 31.10 & 138.00 & Fixed 16 & HLS   & Zynq Z7045 & 125   & 123.12 &       & 219   & 900   & 2400.0 \\
\midrule
\multirow{2}[4]{*}{\cite{Abdelouahab2017}} & SVHN-C & 0.02  & 0.08  & Fixed 5 & \multirow{2}[4]{*}{HDL} & \multirow{2}[4]{*}{Cyclone5 GX} & \multirow{2}[4]{*}{63.96} & 170.73 & \multirow{2}[4]{*}{} & 40    & 0     & 10.9 \\
\cmidrule{2-5}\cmidrule{9-9}\cmidrule{11-13}      & LeNet5-C & 0.04  & 0.03  & Fixed 3 &       &       &       & 2438.46 &       & 8     & 0     & 0.2 \\
\bottomrule
\end{tabular}%

}
\end{table}

    \section{Approximate Computing of CNN Models}
\label{sec:OptAC}
Beside the computational transforms and data-path optimizations, the CNN execution can be accelerated when employing approximate computing which is known to perform efficiently on FPGAs~\cite{Mittal2016}.

In this approach,  a minimal amount of the \ac{cnn} accuracy is traded to improve the computational throughput and energy efficiency of the execution. Two main strategies are employed. This first implements approximate \emph{arithmetic} to process the CNN layers with a reduced precision while the second aims to reduce the number of operations occurring in CNN models without critically affecting the modeling performance. Both of these methods can be integrated in the \emph{learning} phase to jointly maximize the accuracy and minimize the workload of a given CNN model.

\subsection{Approximate Arithmetic for CNNs}
Several studies have demonstrated that the precision of both operations and operands in \acp{cnn}\footnote{and more generally in neural networks} can be reduced without critically affecting their predictive performance. This reduction can be achieved by \textit{quantizing} either or both of the CNN inputs, weights and/or \acp{fm} using a fixed point numerical representation and implementing \textit{approximate multipliers and adders}.

\subsubsection{Fixed point arithmetic:\\}
% \begin{itemize}
% 	\item \Plan{Way more energy efficient then floating point~\cite{David2007}}
% 	\item \Plan{CNNs demonstrate robustness to fixed point arithmetic}
% 	\item \Plan{Courbe: DSP vs nbits, ALMs vs nBits, Accuracy vs nBits}
% 	\item \Plan{Arithmetic representation \ac{sfp} vs \ac{dfp} : Refer to Sze}
%     \item \Plan {\cite{Hubara2016,Courbariaux2014,Zhou2016,Darryl2016,Suyog2015,Anwar2015}}
% 	\item \Plan{Multiple frameworks to train CNNs with fixed point computing: \cite{Gysel2016,Guo2017} for Caffe,\cite{Zhou2016,TensorFlow2017} For TensorFlow}
%     \item \Plan{FPGA Implementations:}
%     \item \Plan{Early works, such~\cite{Chakradhar2010,Farabet2009b}, use \ac{sfp} with a bit-width of 16 bits for FMs and Weights, and a bit-width of 48 bits for the accumulators avoid over-flow}
%     \item \Plan{Accelerator of~\cite{Suda2016} employs different bit-widths for the FMs (6 bits) conv kernels (10 bits) and FC weights(16 bits). Same scheme is used in~\cite{Ma2016,Ma2017}.}
%     \item \Plan{\cite{Gysel2016} investigates \ac{dfp} quantification scheme for CNNs. Used in works \cite{Motamedi2017} to map a CNN accelerator for AlexNet and, in by \cite{Qiu2016} to map VGG. In the latter work, PEs include shift registers to support dynamic quantization.}
% \end{itemize}
In a general way, CNN models are deployed in CPUs and GPUs using the same numerical precision they were trained with, relying on \textit{simple-precision floating point} representation. This format employs 32 bits, arranged according to the IEEE754 standard. In FPGAs, implementations such~\cite{Zhang2015,Alwani2016,Atul2016} employ this data representation.

% as follows:
% \begin{itemize}
% \item A bit $s$ that determines the sign of a given number.
% \item An exponent $e$, coded in 8 bits, that grants a wide numerical range to floating point representation .
% \item A mantissa $m$, coded in 23 bits, that grants good precision to floating point representation.
% \end{itemize}
% \begin{equation}
% 	z_{\text{Float32}}= (-1)^{s} \times 2^{e-127} \times (1+ \frac{m}{23}) 
% \end{equation}
Nonetheless, several studies in~\cite{Anwar2015,Suyog2015,Darryl2016} demonstrate that inference of \acp{cnn} can be achieved with a reduced precision of operands. In addition, works in~\cite{Courbariaux2014,Hubara2016,Zhou2016,Wu2016} demonstrate the applicability of fixed-point arithmetic to train CNNs. In both cases, \acp{fm} and/or weights are \textit{quantized} using a \textit{fixed point representation} scheme.  In simplest version of this format, numbers are encoded with the same bit-width $(bw)$ that is set according to the numerical range and the desired precision. More particularly, all the operands share the same exponent (i.e scale factor) that can be seen as as the position of the radix point. In this paper, we refer to this representation as \ac{sfp}.

% \begin{figure}
% \centering
% \includegraphics[width=0.5\textwidth]{img/SPFQuantizedModels.png}
% \caption{\hl{From~\cite{Suda2016}, JE REDESSINE CA JEUDI SOIR}}
% \label{fig:SPFQuantizedModels}
% \end{figure}

When compared to floating point, \ac{sfp} computing with compact bit-width is known to be more efficient in terms of hardware utilization and power consumption. This is especially true in \acp{fpga}~\cite{David2007}, where a single \ac{dsp} block can either implement \emph{one} 32bits floating point multiplication, \emph{two} $18 \times 19$ bits multiplications, or \emph{three} $18 \times 19$
multiplications~\cite{IntelFPGA2017a}.

This motivated early implementations to employ \ac{sfp} in building FPGA-Based CNN accelerators, such in~\cite{Sankaradas2009,Farabet2009b,Chakradhar2010}, or in~\cite{Farabet2012,Gokhale2014}, where authors use a 16 bits (Q8.8) format to represent FMs and weights. To prevent overflow, the bit-width is expanded when computing the weighted-sums of convolutions and inner-products. If $b_X$ bits are used to quantize the \ac{fm} and $b_{\Theta}$ bits are used to quantize the weights, an accumulator of size $b_{acc}$ is used, according to equation~\ref{eq:accBitwidth}, which corresponds to accumulators of 48 bits in~\cite{Farabet2009b,Chakradhar2010}.
\begin{equation}
\label{eq:accBitwidth}
b_{acc} = b_{x} + b_{\Theta} + \max_{\ell \leq L} \Big( \text{log}_2 \left( C_\ell  K_\ell^2 \right) \Big)
\end{equation}

\subsubsection{Dynamic Fixed Point for CNNs:\\}
In deep topologies, it can be observed that distinct parts of a network can have a significantly different numerical range of data. More particularly, the FMs of deep layers tend to have larger numerical range than first FMs, while the weights are generally much smaller than the FMs. As a consequence, the bit-width is expanded to keep the same precision while preventing overflow, as in\cite{Chakradhar2010}. As a result, and as pointed-out~\cite{Courbariaux2014}, \ac{sfp} with its unique shared fixed exponent, is ill-suited to deep learning.

To address this problem, works in~\cite{Courbariaux2014,Gysel2016} advocates the use of \ac{dfp}~\cite{Williamson1991}\footnote{ An other approach to address this problem is to use half-precision 16 bits floating point, as used in~\cite{Aydonat2017}}. In \ac{dfp}, different scaling factors are used to process different parts of the network. More particularly, weights, weighted sums and outputs of each layer are assigned distinct scale factors. The optimal scale factors and bit-widths (i.e the ones that deliver the best trade-off between accuracy loss and computational load) for each layer can be derived after a brute force exploration using dedicated frameworks that supports \ac{dfp} such~\cite{Gysel2016,Guo2017} for Caffe and~\cite{Zhou2016} for TensorFlow. In addition, these tools can \textit{fine-tune} the CNN model to improve the accuracy of the quantized network. 

The FPGA-Based CNN Accelerator proposed in~\cite{Suda2016} is build upon this quantification scheme and employs different bit-widths to represent the \ac{fm}, the convolution kernels and the FC weights with \textit{resp.} $16$,$8$,$10$ bits. Without fine-tuning, authors report a drop of 1\% in classification accuracy of AlexNet. For the same network, works of~\cite{Ma2016} employs $10$ bits for \acp{fm}, $8$ bits for both \textit{conv} and \textit{FC} weights and report an accuracy drop of 0.4\%. In a same way, Qiu \textit{et al.} employ \ac{dfp} to quantize the VGG with $8$,$8$ and $4$ bits while reporting 2\% of accuracy drop. In these accelerators, dynamic quantization is supported by means of data shift modules~\cite{Qiu2016}. Finally, the accelerator in~\cite{Motamedi2017} rely on the Ristretto framework~\cite{Gysel2016} to derive an AlexNet model wherein the data is quantized in 16 bits with distinct scale factors per layer\footnote{ Since the same PEs are reused to process different layers, the same bit-width is used with a variable radix point for each layer}.

% \begin{figure}
% %\includegraphics[]{}
% \subfloat[]{\includegraphics[width=0.3\textwidth]{img/bitwidth2.pdf}}
% \hfill
% \subfloat[]{\includegraphics[width=0.4\textwidth]{img/bitwidth1.pdf}}
% \caption{\hl{Francois, tu me refais les images stp ? change les nombres}}
% \label{fig:QuantizationSchemes}
% \end{figure}

\subsubsection{Extreme quantification with Binary and pseudo-Binary Nets:\\}
\label{sec:bnn}
\begin{figure}
	\centering
	\subfloat[]{\label{img:bnnLayout}\includegraphics[width=.4\textwidth]{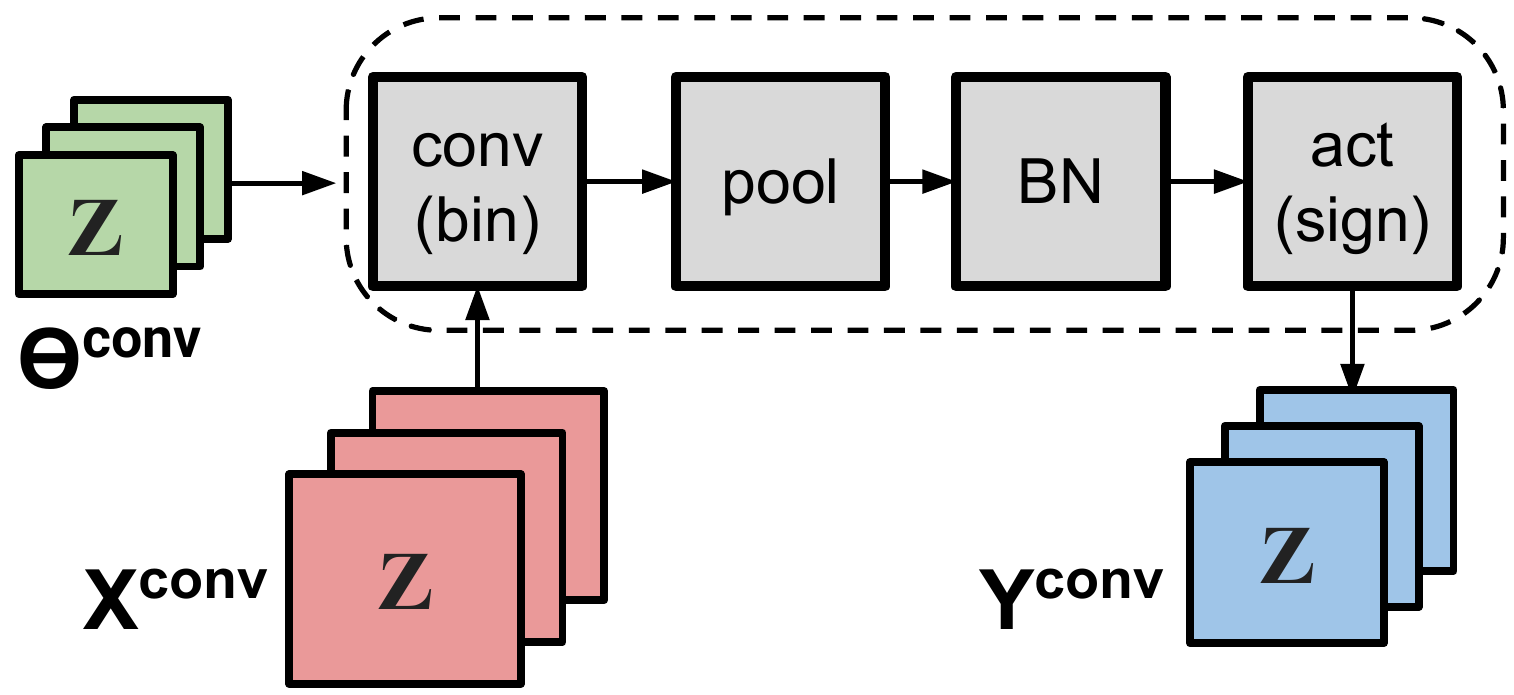}}
    \hfill
	\subfloat[]{\label{img:XNOR}\includegraphics[width=.45\textwidth]{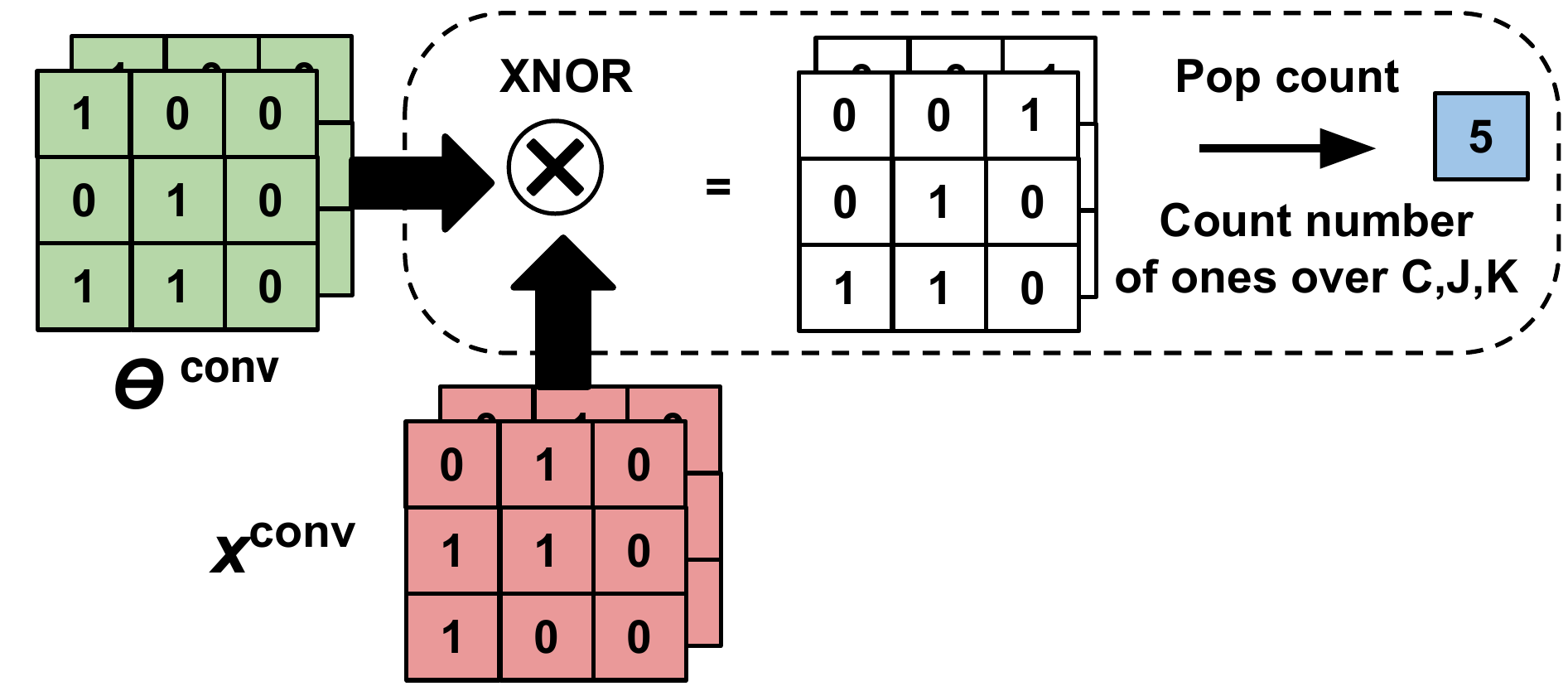}}
	\caption{Binary Neural Networks: a-Processing Pipeline, b-Binary Convolutions}
	\label{img:BNNAll}
\end{figure}

Beside fixed point quantification, training and inferring CNNs with \emph{extremely compact data representations}, is a research area that is gaining interest. In particular, works in BinaryConnect~\cite{Courbariaux2015} investigate the applicability of binary weights (i.e weights with either a value of $-\theta$ or $\theta$) to train \acp{cnn}, which lowers both bandwidth requirements and accuracy on ImageNet by respectively 3200\% and 19.2\% (vs AlexNet Float32 Model).
The same authors go further by implementing \acp{bnn}~\cite{Courbariaux2016}, with a 1bit representation for both \ac{fm} and weights. In these networks, negative data is represented as $0$ while positive values are represented as $1$. As a consequence, the computation of \acp{mac} boils down to an XNOR operation followed by a pop-count, as shown in figure~\ref{img:XNOR}. Moreover, Batch normalization is performed before applying of the $sign$ activation  function in order to reduce the information lost during binarization, as shown in figure~\ref{img:bnnLayout}. However, a classification accuracy drop of 29.8\% is observed on ImageNet when using \acp{bnn}. 
In an attempt to lower the accuracy drop of \acp{bnn}, Rastegari \textit{et al.} proposed XNOR-Nets~\cite{Rastegari2017} which use different scale factors for binary weights (i.e $-\theta_1$ or $+\theta_2$). Moreover, \textit{Pseudo-Binary Networks}, such DoReFa-Net~\cite{Nakahara2017} and QNNs~\cite{Hubara2016c} reduce the accuracy drop to 6.5\% by employing a slightly expanded bit-width (2 bits) to represent the intermediate \acp{fm}.
Finally, in \ac{ttq}~\cite{Zhu2016}, weights are constrained to three values $-\theta_1,0,-\theta_2$ (2 bits), but \ac{fm} are represented in a 32bits float scheme.  As a consequence, the efficiency gain of \ac{ttq} is not as high as in \acp{bnn}.But in turn, \ac{ttq} achieves comparable accuracy on ImageNet, within 0.7\% of full-precision.

In \acp{fpga}, \acp{bnn} benefit from a significant acceleration as the processing of "binary" convolutions can be mapped on XNOR gates followed by a pop count operation, as depicted in figure~\ref{img:XNOR}. Furthermore, and as suggested in~\cite{Nurvitadhi2017}, pop count operation can  be implemented using lookup tables in a way that convolutions are processed only with logical elements.  The \acp{dsp} blocs are can thus be used to process the batch norm calculation (eq~\ref{eq:BatchNorm}, which can be formulated as a linear transform reduces in order reduce the number of operations. This approach is followed in the implementation of~\cite{Zhao2017} to derive an FPGA-Based accelerator for \acp{bnn} that achieves 207.8 GOP/s while only consuming 4.7 W and 3 DSP Blocs to classify the Cifar10 dataset. For the same task, works in~\cite{Umuroglu2017,Fraser2017} use a smaller network configuration\footnote{The network topology used in this work involves 90\% less computations and achieves 7\% less classification accuracy on Cifar10} and reaches a throughput of 2.4 TOP/s when using a larger Zynq 7Z045 Device with 11W Power consumption. For ImageNet classification, Binary Net implementation of~\cite{Liang2017} delivers an overall throughput 1.9 TOP/s on a Stratix V GSD device. In all these works, the first layer is not binerized to achieve better classification accuracy. As pointed-out in~\cite{Liang2017}, the performance in this layer can be improved when using a higher amount of DSP blocs. Finally, an accelerator for ternary neural networks is proposed in~\cite{ProstBoucle2017} and achieves a peak performance of 8.36 TMAC/s at 13W power consumption for Cifar10 Classification.

\subsubsection{Stochastic Computing:\\}
Stochastic Computing (SC) is a low-cost design technique that has been successfully applied in numerous image processing algorithms~\cite{Alaghi2014}. 

In SC, numbers are represented as a random sequence of $s$ bits. In the basic "unipolar" format, the number of ones appearing in the sequence $s$ determines the value of $x$, i.e the numerical value of a given number $x$ is $s_1/s$, where $x$ is the number of \textit{ones} appearing in $s$. The advantage of stochastic  arithmetic is that operations are performed with an ultra-small circuitry. For instance, a single AND gate can map a multiplication. Works in~\cite{Ardakani2015,Ren2016,Kim2016b} demonstrate the feasibility of stochastic arithmetic to accelerate CNNs. More particularly, Ardakani \textit{et al.} propose an FPGA accelerator to classify the MNIST dataset, where multiplications are processed only using AND gates and activation functions (TanH) are implemented in the stochastic domain using FSMs. Such an implementation delivers a computational throughput of 15.44 TOP/s with a misclassification rate of 2.40\% on MNIST. However, one the of weakness of SC are long bit-stream. In fact, to represent an $n$ bits number, a bit-stream $s$ of $2^n$ is required. As a result, stochastic arithmetic suffers from long run-times to perform operations.
Moreover, the generation of this bit-streams resorts to dedicated circuitry known as Stochastic Number Generators (SNGs), which add more overhead to the implementation. As a result, SC-based accelerators implement only shallow neural networks with a limited depth.  
 
\begin{table}
\caption{\ac{fpga}-Based \ac{cnn} accelerators employing Approximate arithmetic}
\label{tab:ApproxArithmetic}
\resizebox{\textwidth}{!}{
% Table generated by Excel2LaTeX from sheet 'ApproxArithmetic'
\begin{tabular}{|c|c|c|c|c|c|c|c|c|c|c|c|c|c|c|c|c|}
\cmidrule{3-17}\multicolumn{1}{c}{\multirow{2}[4]{*}{}} & \multirow{2}[4]{*}{} & \multirow{2}[4]{*}{\textbf{Dataset}} & \multicolumn{2}{c|}{\textbf{Network Workload}} & \multicolumn{4}{c|}{\textbf{Bitwidth}} & \multirow{2}[4]{*}{\textbf{Acc}} & \multirow{2}[4]{*}{\textbf{Device}} & \textbf{Freq} & \textbf{Through.} & \textbf{Power} & \textbf{LUT} & \multirow{2}[4]{*}{\textbf{DSP}} & \textbf{Memory} \\
\cmidrule{4-9}\multicolumn{1}{c}{} &       &       & \textbf{Comp. (GOP)} & \textbf{Param. (M)} & \textbf{In/Out} & \textbf{FMs} & \textbf{W-C} & \textbf{W-FC} &       &       & \textbf{(MHz)} & \textbf{(GOPs)} & \textbf{(W)} & \textbf{(K)} &       & \textbf{(MB)} \\
\midrule
FP32  & \cite{Zhang2017a} & ImageNet & 30.8  & 138.0 & 32    & 32    & 32    & 32    & 90.1  & Arria10 GX1150 & 370   & 866   & 41.7  & 437   & 1320  & 25.0 \\
\midrule
FP16  & \cite{Aydonat2017} & ImageNet & 1.4   & 61.0  & 16    & 16    & 16    & 16    & 79.2  & Arria10 GX1150 & 303   & 1382  & 44.3  & 246   & 1576  & 49.7 \\
\midrule
\multirow{3}[6]{*}{DFP} & \cite{Ma2017} & ImageNet & 30.8  & 138.0 & 16    & 16    & 8     & 8     & 88.1  & Arria10 GX1150 & 150   & 645   &       & 322   & 1518  & 38.0 \\
\cmidrule{2-17}      & \cite{Ma2017a} & ImageNet & 30.8  & 138.0 & 16    & 16    & 16    & 16    &       & Arria10 GX1150 & 200   & 720   &       & 132   & 1518  & 44.5 \\
\cmidrule{2-17}      & \cite{Zhang2017a} & ImageNet & 30.8  & 138.0 & 16    & 16    & 16    & 16    &       & Arria10 GX1150 & 370   & 1790  &       & 437   & 2756  & 29.0 \\
\midrule
\multirow{5}[10]{*}{BNN} & \cite{Zhao2017} & Cifar10 & 1.2   & 13.4  & 20    & 2     & 1     & 1     & 87.7  & Zynq Z7020 & 143   & 208   & 4.7   & 47    & 3     &  \\
\cmidrule{2-17}      & \cite{Umuroglu2017} & Cifar10 & 0.3   & 5.6   & 20/16 & 2     & 1     & 1     & 80.1  & Zynq Z7045 & 200   & 2465  & 11.7  & 83    &       & 7.1 \\
\cmidrule{2-17}      & \multirow{3}[6]{*}{\cite{Liang2017}} & MNIST & 0.0   & 9.6   & 8     & 2     & 1     & 1     & 98.2  & \multirow{3}[6]{*}{Stratix5 GSD8} & \multirow{3}[6]{*}{150} & 5905  & \multirow{3}[6]{*}{26.2} & 364   & 20    & \multirow{3}[6]{*}{44.2} \\
\cmidrule{3-10}\cmidrule{13-13}\cmidrule{15-16}      &       & Cifar10 & 1.2   & 13.4  & 8     & 8     & 1     & 1     & 86.3  &       &       & 9396  &       & 438   & 20    &  \\
\cmidrule{3-10}\cmidrule{13-13}\cmidrule{15-16}      &       & ImageNet & 2.3   & 87.1  & 8     & 32    & 1a    & 1     & 66.8  &       &       & 1964  &       & 462   & 384   &  \\
\midrule
\multirow{3}[6]{*}{TNN} & \multirow{3}[6]{*}{\cite{ProstBoucle2017}} & Cifar10 & 1.2   & 13.4  & \multirow{3}[6]{*}{8} & \multirow{3}[6]{*}{2} & \multirow{3}[6]{*}{2} & \multirow{3}[6]{*}{2} & 89.4  & \multirow{3}[6]{*}{Xilinx7 VX690T} & \multirow{3}[6]{*}{250} & 10962 & 13.6  & 275   &       & 39.4 \\
\cmidrule{3-5}\cmidrule{10-10}\cmidrule{13-17}      &       & SVHN  & 0.3   & 5.6   &       &       &       &       & 97.6  &       &       & 86124 & 7.1   & 155   &       & 12.2 \\
\cmidrule{3-5}\cmidrule{10-10}\cmidrule{13-17}      &       & GTSRB & 0.3   & 5.6   &       &       &       &       & 99.0  &       &       & 86124 & 6.6   & 155   &       & 12.2 \\
\bottomrule
\end{tabular}%

}
\end{table}

\subsection{Reduce Computations in CNNs}
\label{sec:sparcity}
In addition to approximate arithmetic, several studies attempt to the reduce the number of operations involved in CNNs. For FPGA-Based implementation, two main strategies are investigated: \textit{weight pruning}, which increases the \textit{sparsity} of the model, and \textit{low-rank approximation} of filters, which reduces the number of multiplications occurring in the inference.

\subsubsection{Weight Pruning:\\}
As highlighted in~\cite{Liu2015}, \acp{cnn} as over-parametrized networks and a large amount of the weights can be removed --or \textit{pruned}-- without critically affecting the classification accuracy. In its simplest form, pruning is performed according to the magnitude such as the lowest values of the weights are truncated to zero~\cite{Han2015b}. In a more recent approach, weights removal is driven by energy consumption of a given node of the graph, which is 1.74x more efficient than magnitude-based approaches~\cite{Yang2016}. In both approaches, pruning is followed by a fine-tuning of the remaining weights in order to improve the classification accuracy. This is for instance the case in~\cite{Han2016a}, where pruning removes respectively 53\% and 85\% of the weights in AlexNet \textit{conv} and FC layers for less then 0.5\% accuracy loss.

\subsubsection{Low Rank Approximation:\\}
Another way to reduce the computations occurring in CNNs is to maximize the number of of \emph{separable filters} in CNN models. A 2D-separable filter $\theta^{\text{sep}}$ has a unitary rank (i.e $\text{rank} \left(\theta^{\text{sep}} \right) = 1$), and can be expressed as two successive 1D filters $\theta_{J\times 1}$ and $\theta_{1\times K}$. When expanding this to 3D filters, a separable 3D convolution requires $C+J+K$ multiplications while a standard 3D convolution requires $C\times J \times K$ multiplications.

Nonetheless, only a small proportion of the filters in CNN Models are separable. To increase this proportion, a first approach is to force the convolution kernels to be separable by penalizing \textit{high rank filters} when training the network~\cite{Sironi2015}. Alternatively, and after the training, the weights $\matr{\Theta}$ of a given layer can be approximated into a small set of $r$ \textit{low rank filters} that can be implemented as a succession of fully separable filters. In this case, $r\times(C+J+K)$ multiplications are required to process a single 3D-convolution. 

For FC layers, in which the processing boils down to a vector-matrix product, low rank approximation can be achieved by employing, for instance, the SVD decomposition of the weight matrix $\matr{\tilde{\Theta}}^{\text{fc}}$ (cf. sec~\ref{sec:GEMM-FC}). Finally, and in a same way to pruning, low rank approximation of weights is followed by a fine-tuning in order counterbalance the classification accuracy drop.

\subsubsection{\ac{fpga} Implementations:\\}
In FPGA Implementations, low rank approximation is applied on FC layer to significantly reduce the number of weight, such as in~\cite{Qiu2016}, where authors derive a VGG16-SVD model that achieves 87.96\% accuracy on ImageNet with 63\% less parameters. 

Sparsity in pruned \acp{cnn} can be exploited in \ac{fpga} implementations by fully unrolling the processing of a given layer, and skipping (i.e not mapping) the multiplications with zero weights. This approach is investigated in~\cite{Abdelouahab2017}, but can be infeasible when the resource of a given device doesn't match with computational requirements of a given layer.  
Instead, sparsity and pruning can be exploited when processing \textit{conv} and \textit{fc} layers as \ac{gemm} (c.f~\ref{sec:GEMM}. In this case, the challenge is to determine the optimal format of matrices that maximizes the chance to detect and skip zero computations, such compressed sparse column (CRC) or compressed sparse row (CSR) formats\footnote{These format represents a matrix by three one-dimensional arrays, that respectively contain nonzero values, row indices and column indices}.  Based on previous studied related to sparse \ac{gemm} implementation on \acp{fpga} in~\cite{Dorrance2014}, Sze \textit{et al.}\cite{Sze2017} advocates the use of the CRC to process \acp{cnn} because this format provides a lower memory bandwidth when the output matrix is smaller then the input, which is typically the case in \acp{cnn} where $N < CJK$ in Fig~\ref{fig:GEMM-Conv}. 

However, this efficiency of CRC format is only valid for extremely sparse matrices (typically with $\le 1$\% of non zeros), while pruned \ac{cnn} matrices are not that sparse (typically, $\le 4-80$\% of non zeros). Therefore, works in~\cite{Nurvitadhi2017} use a \textit{zero skip scheduler}, which is an on-chip data manager thanks to which zero elements are identified and not scheduled onto the \ac{mac} processing. As a result, the number of cycles required to compute the sparse \ac{gemm} is reduced, which corresponds to a 4x speedup in cycle count for and 85\% sparse AlexNet layers. Finally, authors report to a projected throughput of 12 TOP/s for pruned \acp{cnn} in the next Intel Stratix10 FPGAs, which outperforms and the computational throughput of state-of-the-art \ac{gpu} implementations by 10\%.   

\begin{table}
\caption{\ac{fpga}-Based \ac{cnn} accelerators employing pruning and low rank approximation}
\label{tab:CompReduction}
\resizebox{\textwidth}{!}{
% Table generated by Excel2LaTeX from sheet 'CompReduction'
\begin{tabular}{|c|c|c|c|c|c|c|c|c|c|c|c|c|c|c|}
\cmidrule{3-15}\multicolumn{1}{c}{\multirow{2}[4]{*}{}} & \multirow{2}[4]{*}{} & \multirow{2}[4]{*}{\textbf{Dataset}} & \multicolumn{2}{c|}{\textbf{Network Workload}} & \textbf{Removed} & \multirow{2}[4]{*}{\textbf{Bitwidth}} & \textbf{Acc} & \multirow{2}[4]{*}{\textbf{Device}} & \textbf{Freq} & \textbf{Through.} & \textbf{Power} & \textbf{LUT} & \multirow{2}[4]{*}{\textbf{DSP}} & \textbf{Memory} \\
\cmidrule{4-5}\multicolumn{1}{c}{} &       &       & \textbf{Comp. (GOP)} & \textbf{Param. (M)} & \textbf{ Param. (\%)} &       & \textbf{(\%)} &       & \textbf{(MHz)} & \textbf{(GOPs)} & \textbf{(W)} & \textbf{(K)} &       & \textbf{(MB)} \\
\midrule
SVD   & \cite{Qiu2016} & ImageNet & 30.5  & 138.0 & 63.6  & 16 Fixed & 87.96 & Zynq 7Z045 & 150   & 137   & 9.6   & 183   & 780   & 17.5 \\
\midrule
\multirow{2}[4]{*}{Pruning} & \cite{Fujii2017} & Cifar10 & 0.3   & 132.9 & 89.3  & 8 Fixed & 91.53 & Kintex 7K325T & 100   & 8621  & 7.0   & 17    & 145   & 15.1 \\
\cmidrule{2-15}      & \cite{Nurvitadhi2017} & ImageNet & 1.4   & 61.0  & 85.0  & 32 Float & 79.70 & Stratix 10 & 500   & 12000 & 141.2 &       &       &  \\
\bottomrule
\end{tabular}%

}
\end{table}

    \section{Conclusion}
\label{sec:Conclusion}

In this paper, a number of methods and tools have been compared that aim at porting Convolutional Neural Networks onto FPGAs. At the network level, approximate computing and datapath optimization methods have been covered while at the neuron level, the optimizations of convolutional and fully connected layers have been detailed and compared. All the different degrees of freedom offered by FPGAs (custom data types, local data streams, dedicated processors, etc.) are exploited by the presented methods. Moreover, algorithmic and datapath optimizations can a should be jointly implemented, resulting in additive hardware performance gains.

CNNs are by nature overparameterized and support particularly well approximate computing techniques such as weight pruning and fixed point computation. Approximate computing already constitutes a key to CNN acceleration over hardware and will certainly continue driving the performance gains in the years to come.

%\bibliographystyle{ACM-Reference-Format}
%\bibliography{libraryCompact}
\bibliographystyle{unsrt}
\bibliography{Mendeley}
\end{document}